\documentclass[notitlepage,nofootinbib]{revtex4-1}

\usepackage{amsmath}
\usepackage{amsfonts,amssymb}
\usepackage{hyperref}
\usepackage{graphicx}
\usepackage{natbib}
\usepackage[utf8]{inputenc}
\usepackage{xcolor}
\usepackage{bm}
\usepackage{multirow}
\usepackage{diagbox}
\usepackage{hhline}
\usepackage[normalem]{ulem}

\usepackage[normalem]{ulem}

\usepackage{xcolor}

\begin{document}

\title{An effective description of Laniakea: impact on cosmology and the local determination of the Hubble constant}

\author{Leonardo Giani$^a$}
\email{l.giani@uq.edu.au}
\affiliation{$^a$ School of Mathematics and Physics, The University of Queensland, Brisbane, QLD 4072, Australia}

 \author{Cullan Howlett$^a$}

 \affiliation{$^a$ School of Mathematics and Physics, The University of Queensland, Brisbane, QLD 4072, Australia}

 \author{Khaled Said$^a$}
 \affiliation{$^a$ School of Mathematics and Physics, The University of Queensland, Brisbane, QLD 4072, Australia}

 \author{Tamara Davis$^a$}

 \affiliation{$^a$ School of Mathematics and Physics, The University of Queensland, Brisbane, QLD 4072, Australia}

 \author{Sunny Vagnozzi$^{b\;c}$}

 \affiliation{$^b$ Department of Physics, University of Trento, Via Sommarive 14, 38123 Povo (TN), Italy}
  \affiliation{$^c$ Trento Institute for Fundamental Physics and Applications (TIFPA)-INFN, Via Sommarive 14, 38123 Povo (TN), Italy}

\begin{abstract}
\noindent We propose an effective model to describe the bias induced on cosmological observables by Laniakea, the gravitational supercluster hosting the Milky Way, which was defined using peculiar velocity data from Cosmicflows-4 (CF4). The structure is well described by an ellipsoidal shape exhibiting triaxial expansion, reasonably approximated by a constant expansion rate along the principal axes. Our best fits suggest that the ellipsoid, after subtracting the background expansion, contracts along the two smaller axes and expands along the longest one, predicting an average expansion of $\sim -1.7 ~\rm{km}/\rm{s}/\rm{Mpc}$. The different expansion rates within the region, relative to the mean cosmological expansion, induce line-of-sight-dependent corrections in the computation of luminosity distances. We apply these corrections to two low-redshift datasets: the Pantheon+ catalog of type Ia Supernovae (SN~Ia), and 63 measurements of Surface Brightness Fluctuations (SBF) of early-type massive galaxies from the MASSIVE survey. We find corrections on 
 the distances of order $\sim 2-3\%$, resulting in a shift in the inferred best-fit values of the Hubble constant $H_0$ of order $\Delta H_0^{\rm{SN~Ia}}\approx 0.6 ~\rm{km}/\rm{s}/\rm{Mpc}$ and $\Delta H_0^{\rm{SBF}}\approx 1.4 ~\rm{km}/\rm{s}/\rm{Mpc}$, seemingly worsening the Hubble tension.        
\end{abstract}

\maketitle

\section{Introduction}
Roughly a century ago, astronomers collected evidence that we live in an expanding Universe~\cite{1929PNAS...15..168H}. To date, the most commonly accepted description of such an expansion is built upon the assumptions of homogeneity and isotropy of the Universe on sufficiently large scales, encoded in a class of solutions of the Einstein field equations based on the Friedmann-Lema\^itre-Robertson-Walker (FLRW) class of metrics:
\begin{equation}\label{FLRWmetric}
    ds^2 = -c^2dt^2 + a^2(t)\left(\frac{dr^2}{1-Kr^2} + r^2 d\Omega^2\right)\;,
\end{equation}
where $K$ is the (constant) spatial curvature, and $a(t)$ is the scale factor. Null geodesics for the above metric indicate that the spectra of photons travelling in an expanding Universe shift towards the red. About two decades ago, observations of photons produced by the explosions of type Ia supernovae (SN~Ia) provided evidence that the expansion of the Universe is accelerating~\cite{SupernovaCosmologyProject:1998vns,SupernovaSearchTeam:1998fmf}, challenging our understanding of gravity as a solely attractive force. Within the frameworks of General Relativity and FLRW models, such an acceleration requires an exotic source of energy-momentum labelled Dark Energy (DE). The standard cosmological model, the $\Lambda$CDM model (also dubbed concordance model), describes a spatially flat ($K=0$) FLRW Universe in which DE is interpreted as a cosmological constant $\Lambda$~\cite{Carroll:2000fy}, and where most of the matter in the Universe is described by a collisionless, non-relativistic perfect fluid labelled Cold Dark Matter (CDM)~\cite{Profumo:2019ujg}. 

The success and popularity of the concordance model comes from its effectiveness at describing a variety of cosmological observations with only 6 parameters~\cite{DES:2021wwk,eBOSS:2020yzd,Mossa:2020gjc,Brout:2022vxf,Planck:2018vyg}.  
On the other hand, their inference heavily relies on the existence of sources with standardized physical properties, which allow the computation of distances and the construction, after assuming that the law of physics are the same everywhere, of a cosmic ``distance ladder''. To some extent, this jeopardizes the predictive power of the $\Lambda$CDM model since some of the hypotheses that we would like to verify through cosmological observations are sometimes assumed to interpret the observations themselves. To phrase it differently, it is not always possible to derive distances to standardized sources without assuming a cosmological model, which inevitably introduces biases in our cosmological inferences. One such bias could be our underlying assumptions of homogeneity and isotropy, i.e.\ the Cosmological Principle (CP). Whilst the latter seems a reasonable working assumption \textit{a posteriori}, supported by a variety of uncorrelated observations~\cite{Planck:2018vyg,WMAP:2012fli,Scrimgeour:2012wt,Ntelis:2017nrj,Laurent:2016eqo}, the existence of (large-scale) structures reveals the limits of the CP~\cite{Singal:2014wra,Bengaly:2017slg,Migkas:2020fza,Secrest:2020has,Nadolny:2021hti,Secrest:2022uvx,Aluri:2022hzs,Whitford:2023oww}. 

To deal with these limitations, a particularly successful approach is to consider small perturbations on top of a spatially flat FLRW model. These small perturbations are well understood in the linear regime, where the Einstein field equations can be solved analytically, and provide a powerful tool to describe the evolution of the early Universe. Eventually,  the small perturbations that are not smoothed out by the cosmological expansion survive long enough to enter the non-linear regime, and evolve into the complex cosmic web of structures we observe today. Unfortunately, the non-linearity of the Einstein field equations makes it difficult to describe the spacetime geometry around these structures, and analytical solutions can be obtained only for simplified models with a high degree of symmetry.

It is worth stressing that we perform our cosmological inference in a highly inhomogeneous and anisotropic environment, and therefore we are far from being Copernican observers. The impact an inhomogeneous environment could have on the observer and on the background geometry has been extensively discussed in the literature~\cite{1963SvA.....6..699S,Ellis:1987zz,Zalaletdinov:1992cg,Wiltshire:2011vy,Clarkson:2011zq,Buchert:2011sx,Bolejko:2016qku,Koksbang:2019cen,Koksbang:2021qqc,Schander:2021pgt}, and is usually referred to as \textit{backreaction}. The effects of the latter are often quantified by means of some averaging scheme~\cite{Coley:2005ei,Marra:2007pm,Bolejko:2017lai,Koksbang:2019glb,Buchert:2019mvq}, within the working hypothesis that the averaged geometry on sufficiently large scales is well described by the FLRW metric, and that on intermediate scales backreaction is described by some effective modification of the (averaged) Friedmann equations. Unfortunately, at the time of writing there is no general consensus for what concerns how to perform these averages, and their impact on cosmological observables, with a spectrum of possibilities ranging from no impact at all ~\cite{Ishibashi:2005sj,Green:2014aga,Macpherson:2018btl}, to backreaction being responsible for the inferred accelerated expansion of the Universe ~\cite{Wiltshire:2009db,Heinesen:2020sre} (see also Refs.~\cite{Fleury:2013sna,Bolejko:2015gmk} for examples of the intermediate spectrum, and Ref.~\cite{Kolb:2009rp} for a proposed classification of different backreaction mechanisms).

With current surveys reaching $\sim {\cal O}(1\%)$ accuracy in the determination of cosmological parameters, and the concurrent appearance of tensions among independent inferences of the latter~\cite{Planck:2018vyg,Riess:2021jrx,DES:2021wwk}, it is becoming of utmost importance to assess the impact of the local environment on our cosmological inference. 
It is possible to adopt a completely agnostic attitude for what concerns the details of such an environment and adopt the cosmographic approach recently developed and used in Refs.~\cite{Heinesen:2020bej,Heinesen:2020pms,Macpherson:2021gbh,Heinesen:2021azp,Dhawan:2022lze}. Despite being a flexible and powerful framework, the latter comes at the cost of a large and potentially impractical number of free parameters.  Another option is to start from a few simplifying assumptions about the nature of the inhomogeneities, and build upon these to compute their impact on cosmology by directly solving the Einstein field equations. This approach was recently adopted, among others, in Refs.~\cite{Kenworthy:2019qwq,Camarena:2021mjr,Marra:2022ixf,Camarena:2022iae}, where the Universe was described using the Lema\^itre-Tolmann-Bondi (LTB) metric~\cite{Mukhanov:2005sc,Enqvist:2007vb}, a spherically symmetric inhomogeneous solution of the Einstein field equation which recovers (i.e.\ can be made indistinguishable from) a flat FLRW Universe on cosmological scales.
 
In this paper we shall adopt a similar approach, making a few simplifying symmetry assumptions for what concerns the geometrical nature of our local environment. Specifically, our goal is to derive an effective model for our \textit{home in the Cosmos}, Laniakea~\cite{Tully:2014gfa,Dupuy:2023ffz}, a large supercluster hosting the Milky Way. The tempting assumption of spherical symmetry appears unfeasible for such a supercluster, and we therefore seek a description capturing the impact of its anisotropies. In the remainder of this paper we argue that the latter can be, at first order, well described by an ellipsoidal inhomogeneity exhibiting triaxial expansion, moving with peculiar velocity $v_{\rm BF}$ with respect to the cosmological rest frame, where $v_{\rm BF}$ is the bulk flow (or average peculiar velocity) within the inhomogeneity.  We will assume that the metric of the Universe can be effectively described by a piece-wise solution, such that the interior of Laniakea is spatially flat and characterized by three different scale factors, and where the exterior part is described by the standard flat FLRW background. We stress that such a description can at best be an effective one, as it introduces discontinuities in the metric at the boundary of the ellipsoid. Using our effective description we compute the impact of the different expansion rates in the interior of Laniakea on the redshift-distance relation for an observer located in the Milky Way. It is important to stress that these corrections do not alter the background geometry and depend on the position of the observer, and therefore can be classified as \textit{weak backreaction} effects, according to the nomenclature proposed in Ref.~\cite{Kolb:2009rp}.

The rest of the paper is then organized as follows. In Sec.~\ref{Laniakea} we model Laniakea using the Cosmicflows-4 (CF4) velocity field reconstruction. In Sec.~\ref{Backreaction_cosmo} we derive the corrections induced by the effective model for Laniakea on comoving distances and redshifts, and the possible effects thereof on cosmological parameter inference, with particular attention to the Hubble constant $H_0$. Finally, in Sec.~\ref{conclusion}, we summarize our findings, discuss our results, and draw concluding remarks. A more technical discussion concerning luminosity distances in our toy model for Laniakea versus the FLRW metric can be found in Appendix~\ref{appA}.

\section{A Toy model for Laniakea}

\label{Laniakea}

Our goal is to derive an effective description of Laniakea going beyond the traditional working assumption of spherical symmetry to model the local Universe.\footnote{The code used to derive the main results of this paper is publicly available at \href{https://github.com/Leolardo/Laniakea_Backreaction_nb}{https://github.com/Leolardo/Laniakea\_Backreaction\_nb}.} To get a qualitative understanding of the influence of anisotropies, we propose a toy model which describes Laniakea as a homogeneous ellipsoid, expanding at different rates along its principal axes. Being homogeneous, the interior of the ellipsoid is assumed to be at rest in its comoving frame. In other words, any relative motion between a pair of points in the interior is due to the triaxial expansion, and there are no peculiar motions induced by density gradients.
With such a description, the interior of the ellipsoid behaves similarly to a Bianchi I spacetime filled with dust and a cosmological constant, as described by the generalized Heckmann-Schucking solution~\cite{Kamenshchik:2009dt}. Overdensities exhibiting triaxial profiles naturally emerge from ellipsoidal gravitational collapse~\cite{Sheth:1999su,Angrick:2010qg}, and a Bianchi I profile in particular is expected when the spatial curvature of the inhomogeneity is negligible during the linear regime of a fully anisotropic gravitational collapse~\cite{Giani:2021gbs,Giani:2022wda}. 
However, in the following we neglect any time evolution for the anisotropies and assume instead that inside Laniakea the background Hubble factor is shifted by a line-of-sight dependent constant. In other words, given the fact that Laniakea's effective redshift is extremely low, we can safely assume that corrections induced by the time evolution of the anisotropies are at first order negligible. 


\subsection{What is Laniakea?}
Peculiar velocities of galaxies, interpreted as the difference between their recessional velocities caused by the expansion of the Universe and their observed velocities (also corrected for the observer's peculiar motion), can be used to partition the inhomogeneous local Universe in gravitational basins of attraction (see e.g.\ Refs.~\cite{Dupuy:2019blz} for the detailed definition of the latter). These basins are obtained by studying the behaviour of streamlines, that is, curves tangent to the peculiar velocity field. To construct a streamline one starts from an arbitrary initial point, labelled \textit{seed}, and integrates its peculiar velocity field for an infinitesimal displacement (i.e. to a nearby cell). The curve connecting the two tangent vectors at these points defines the beginning of the streamline. One then repeats the process iteratively, until the peculiar velocity field 
eventually reaches a critical point, i.e. an attractor, where the streamline stops. A basin of attraction is defined as a spatial region such that all the streamlines therein contained converge towards the same attractor. In other words, a basin of attraction can be interpreted as a region of space where, apart from the cosmological expansion, the trajectories of test particles are mainly influenced by the gravitational field within its volume. The volumes associated to these gravitational basins are therefore used to identify large scale structures in the local Universe.

In Ref.~\cite{Tully:2014gfa}, using velocity measurements from Cosmicflows-2 (CF2)~\cite{Tully:2013wqa} to reconstruct the (linear) peculiar velocity and density fields in the local Universe, Laniakea was defined as one such gravitational basin and recognized as being the supercluster containing the Milky Way.  In this work we shall consider the most recent determination of Laniakea obtained in Ref.~\cite{Dupuy:2023ffz}, which uses the (grouped) velocity and density field reconstruction of Ref.~\cite{Courtois:2022mxo}, derived from the latest Cosmicflows-4 (CF4) data release~\cite{Tully:2022rbj}. In this reconstruction, Laniakea consists of 4079 cells within a grid of $128^3$ cells covering a volume of $1 \left(\rm{Gpc}/h\right)^3$, with $h$ being the reduced Hubble constant. For reference, the furthest and closest cells on its boundary are at redshifts $z=0.082$ and $z=0.005$ respectively. Our catalog for the velocity field consists of a $64^3$ cells grid covering the same volume with a lower resolution.\footnote{Both grids are publicly available at \href{https://projets.ip2i.in2p3.fr//cosmicflows/}{https://projets.ip2i.in2p3.fr//cosmicflows/}.} The velocities and dispersions for the cells in Laniakea are therefore obtained from the lower resolution grid using a first order cloud-in-cell interpolation. 

It is interesting to assess the likelihood of a structure like Laniakea in the $\Lambda$CDM model. The velocity field reconstruction of Refs.~\cite{Dupuy:2023ffz,Courtois:2022mxo} predicts that the average density contrast in the interior of Laniakea is $\langle \delta \rangle = 0.012$. We can compare this value with $\sigma_R$, the expected variance in the density field predicted by the $\Lambda$CDM model\footnote{For this calculation we assume a flat $\Lambda$CDM model with $\Omega_m = 0.31$} for a sphere of radius $R$. For a sphere covering the same volume as Laniakea, i.e. with radius $R = 110$ Mpc$/h$, we found $\sigma_{110/h} = 0.058$, implying that these types of structures are well within the predictions of the standard model.


\subsection{Spatial Geometry}

\label{bestfitellipsoid}

The starting point for our analysis is to derive an effective description of the spatial geometry of Laniakea. To capture its anisotropies, we look for the ellipsoidal shape which best describes  the $4079$ cells defining the structure. 
Our best fit\footnote{We use a python package publicly available at \href{https://github.com/aleksandrbazhin/ellipsoid_fit_python}{https://github.com/aleksandrbazhin/ellipsoid\_fit\_python}.} is  an ellipsoid centered at supergalactic coordinates SGX $= -9.23$, SGY $= -47.06$, and SGZ $=85.08$, with semiaxes of lengths $48.61$, $128.74$ and $166.38$ in units of Mpc$/h$.
This ellipsoid is rotated with respect to the supergalactic Cartesian axes SGX, SGY, and SGZ by angles of $ -27^\circ$, $ 16^\circ$, and $74^\circ$ respectively. 

\begin{figure}[!tbp]
    \centering
    \includegraphics[scale=0.60]{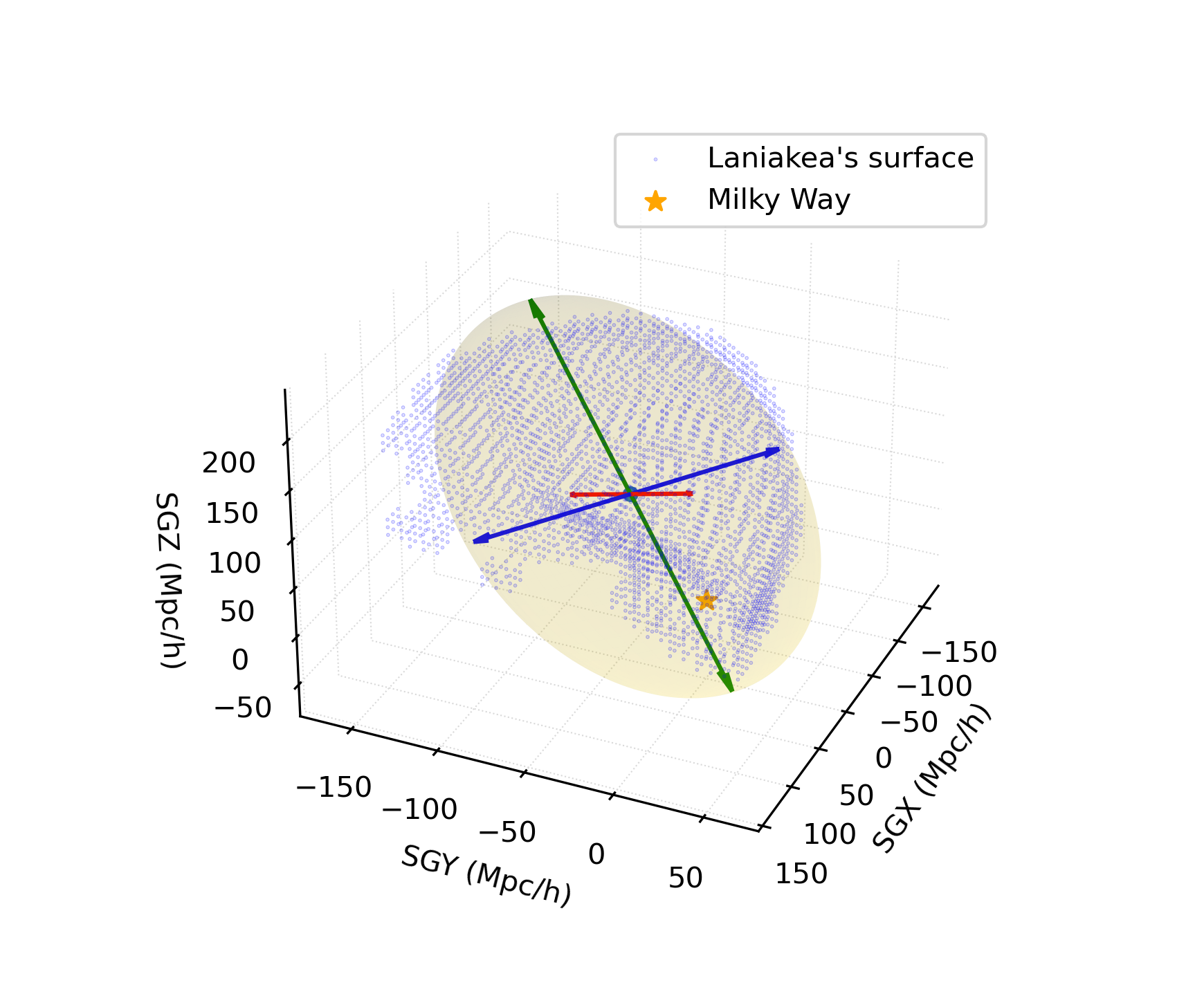}\includegraphics[scale=0.60]{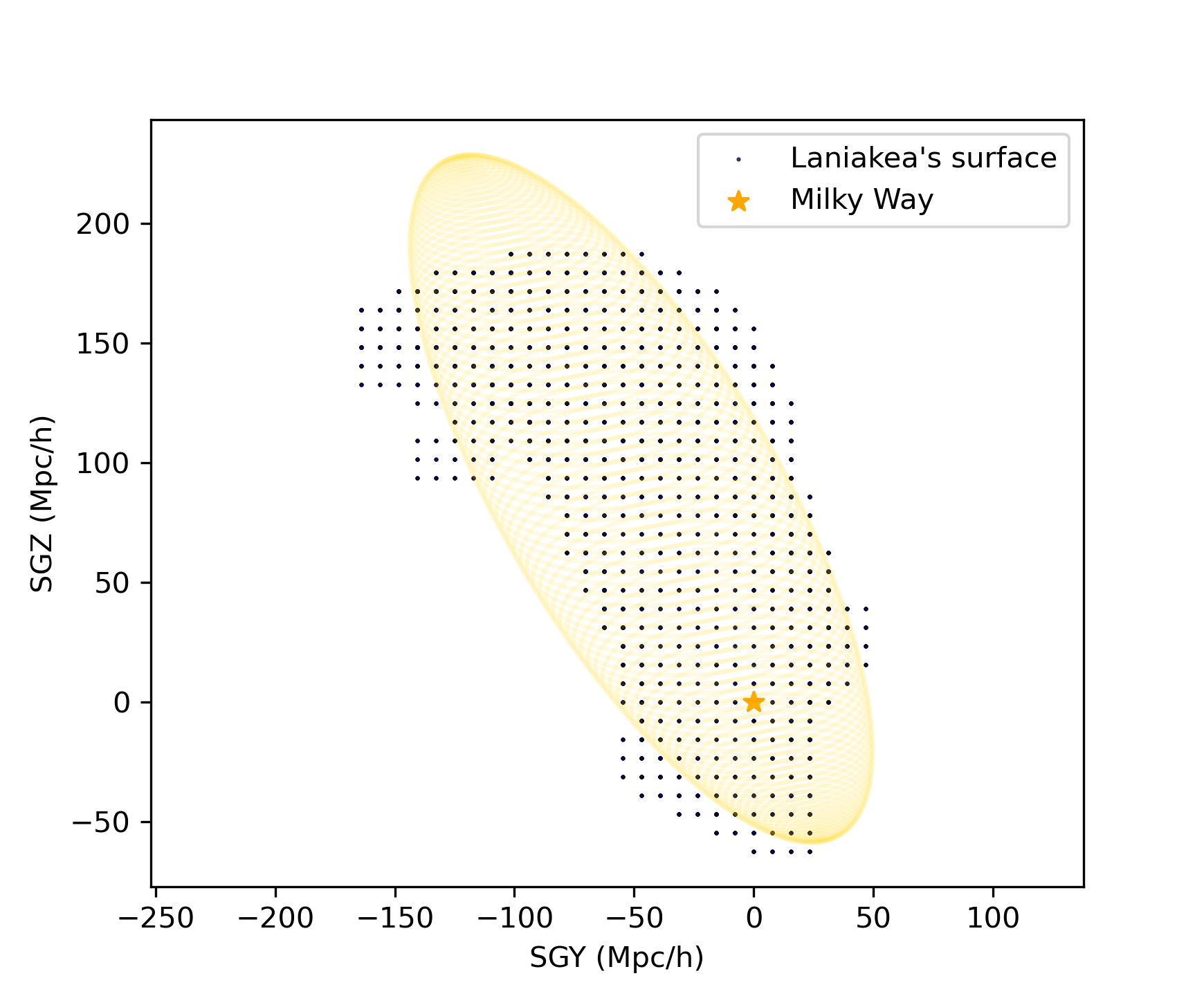}\\
 \includegraphics[scale=0.60]{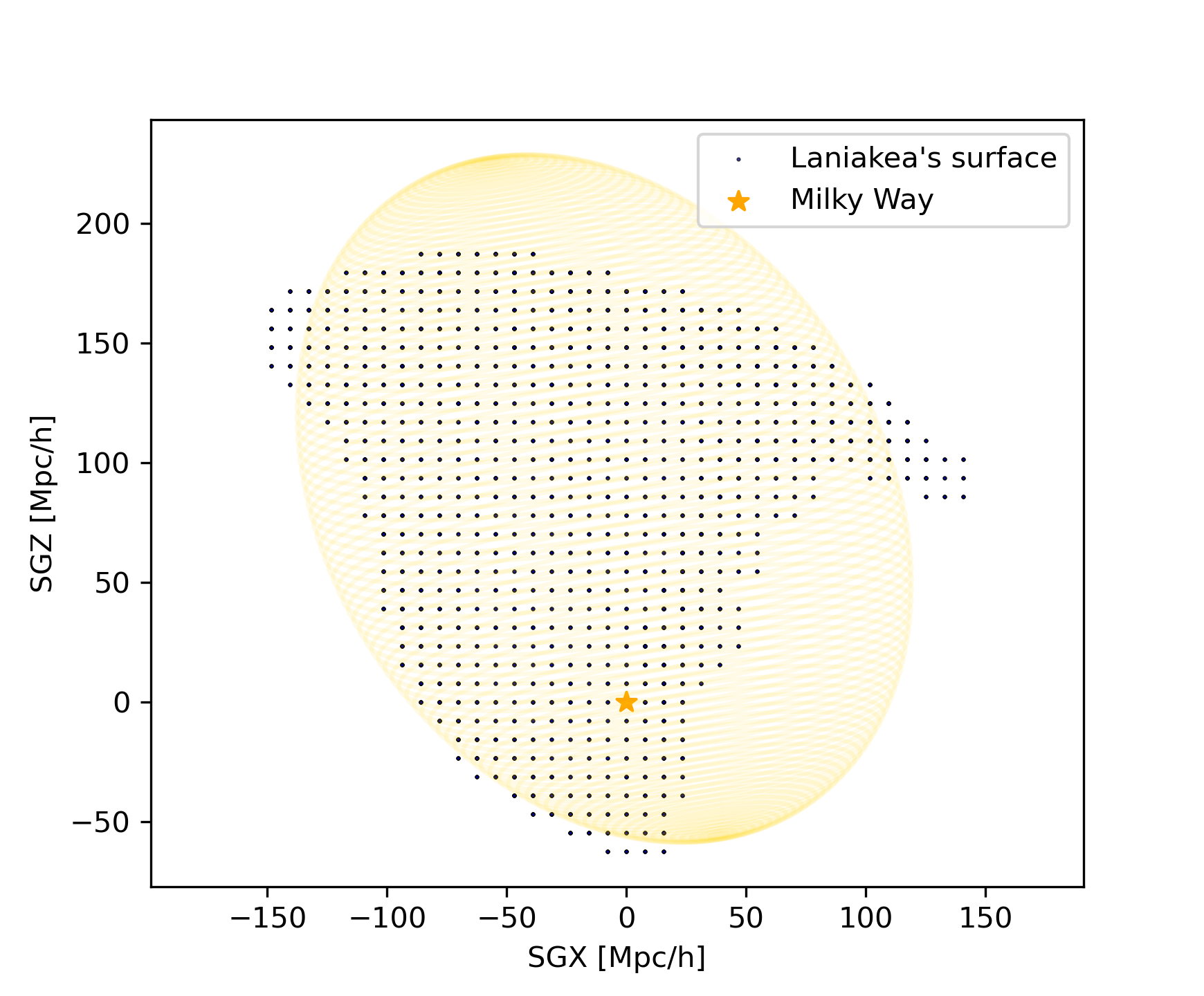}\includegraphics[scale=0.60]{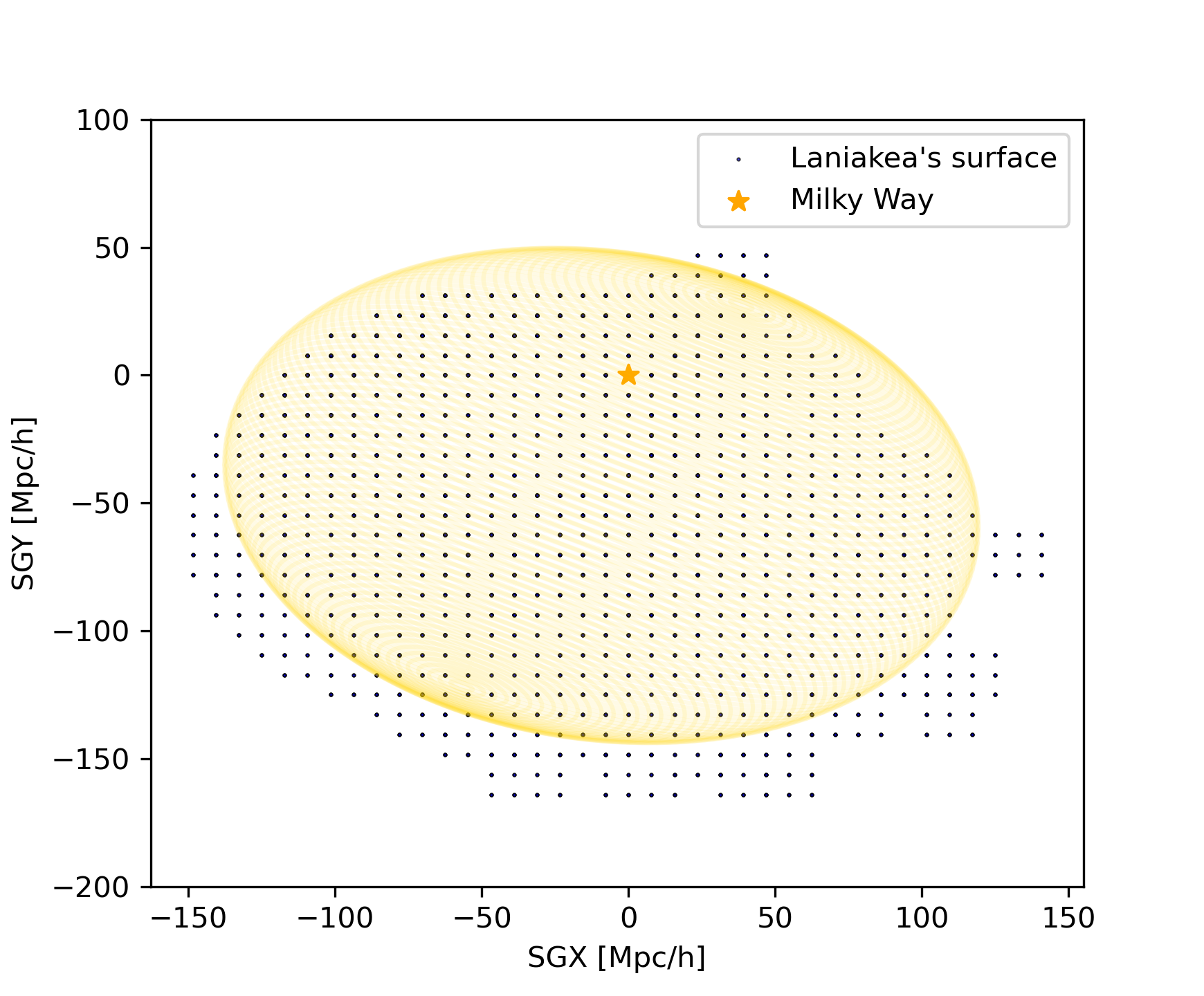}\\
	\caption{Projection along orthogonal planes in the supergalactic Cartesian frame of the cells constituting Laniakea's surface in blue, with overlapped in light yellow the ellipsoid which best fits their volume distribution. The orange star indicates the position of the Milky Way.}
	\label{Ellipsfit}
\end{figure}

The goodness of the fit can be tested through a conformal rescaling of the axes of the ellipsoid into a unit sphere and by computing the average radius and standard deviation thereof. Doing this exercise, we find an average radius of $0.97$ and a standard deviation of $0.23$, indicating that the ellipsoid is a good description on average, but that large spatial deviations occur. For reference, in Fig.~\ref{Ellipsfit} we plot the surface of Laniakea alongside our best-fit ellipsoid, its centre and principal axes.

\subsection{Laniakea comoving frame and peculiar velocities}

It is convenient to work with the peculiar velocity field in the Laniakea Comoving Frame (LCF). The latter is defined as the frame in which the interior of Laniakea is (on average) at rest, with origin at the centre of the ellipsoid and Cartesian coordinates defined by the principal axes. 

\begin{figure}[!tbp]
    \centering
	\includegraphics[scale=0.6]{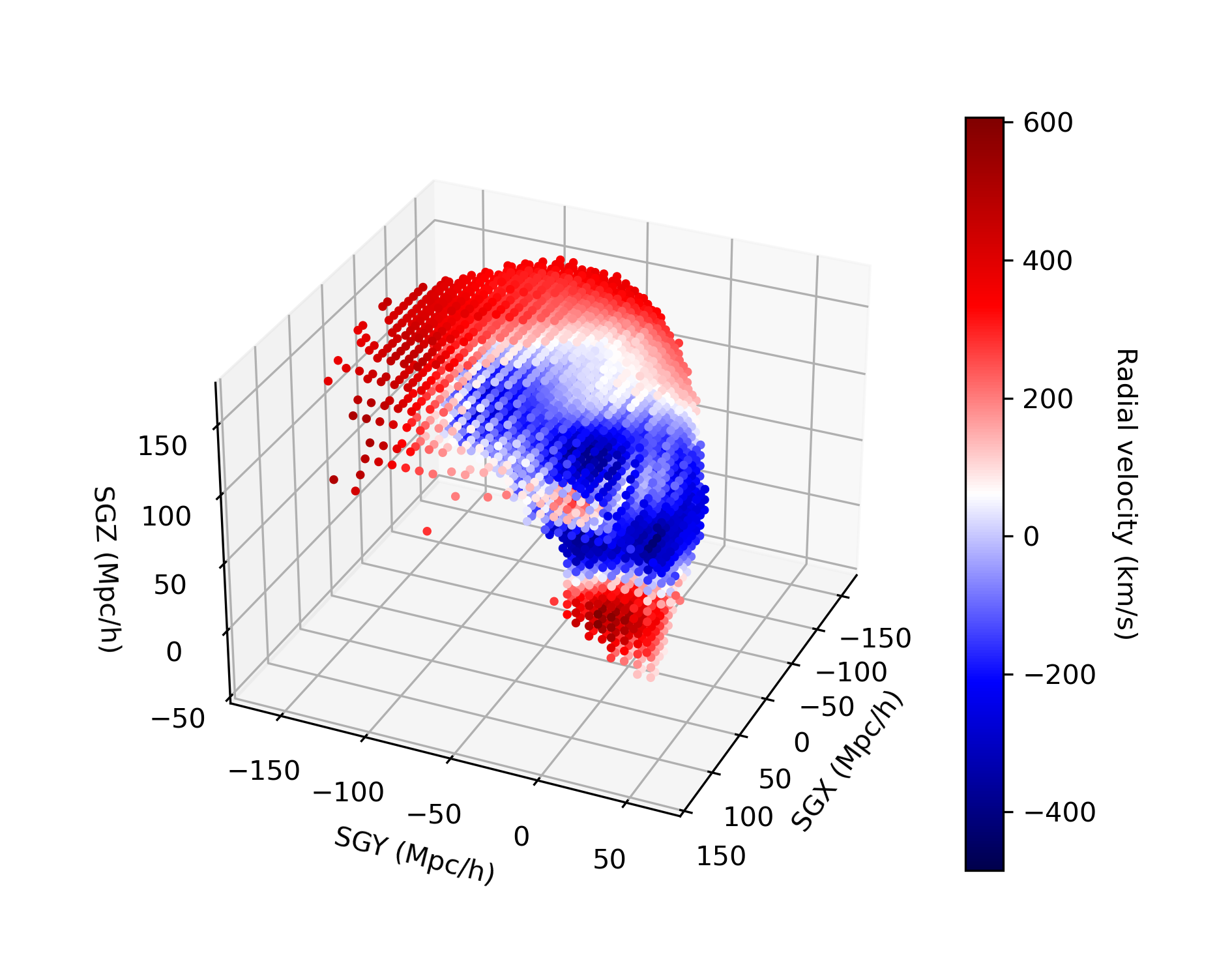}\includegraphics[scale=0.6]{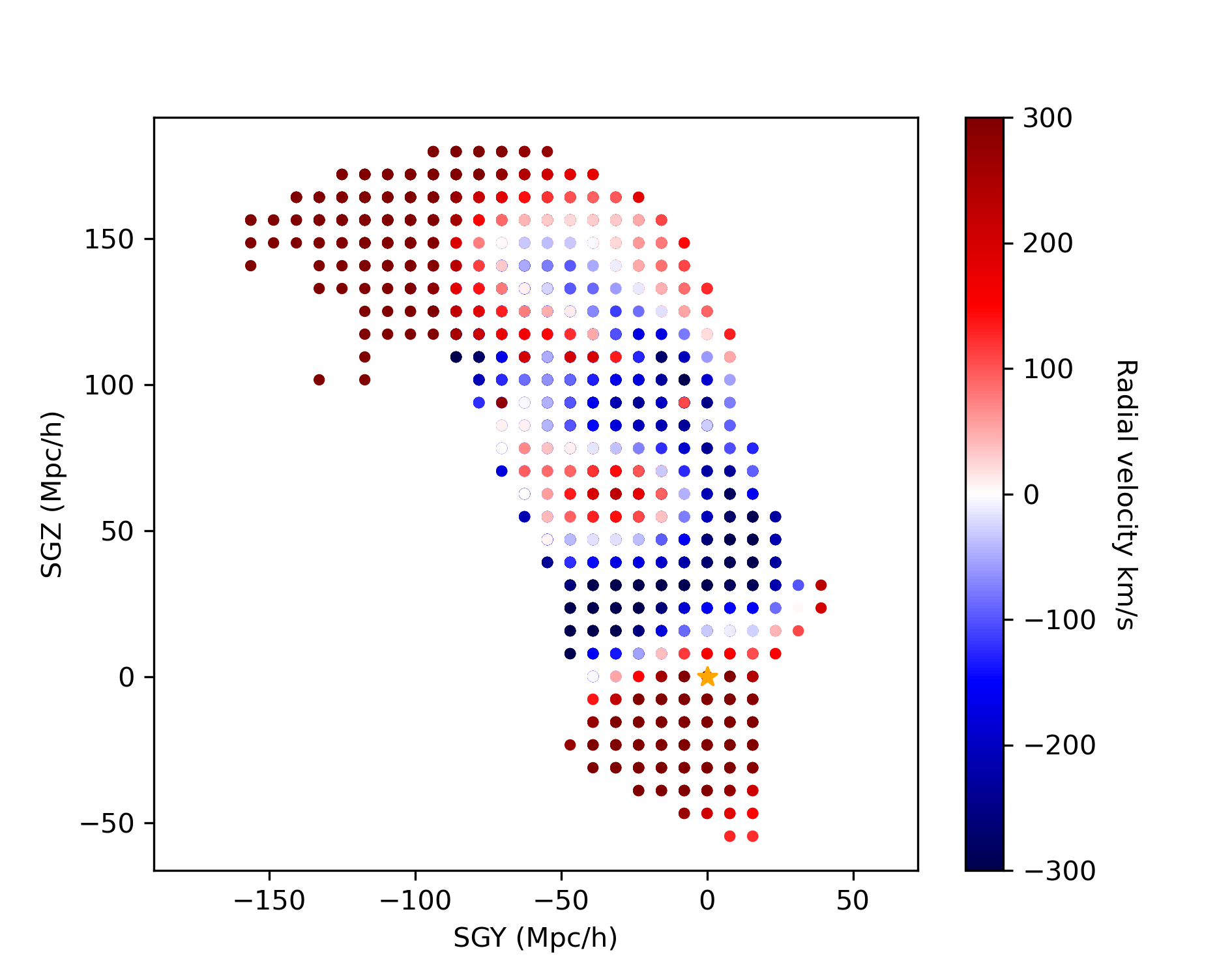} \includegraphics[scale=0.6]{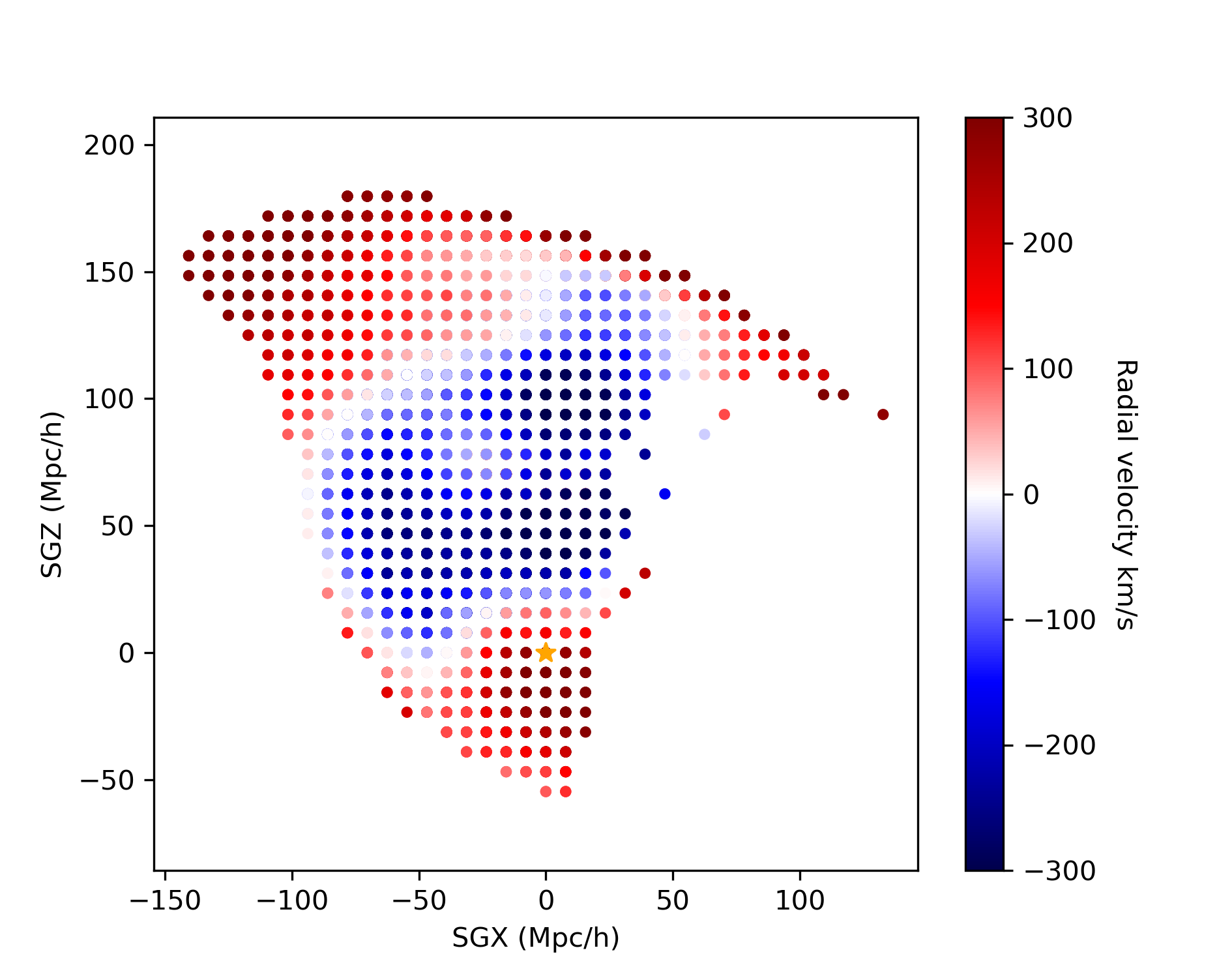}\includegraphics[scale=0.6]{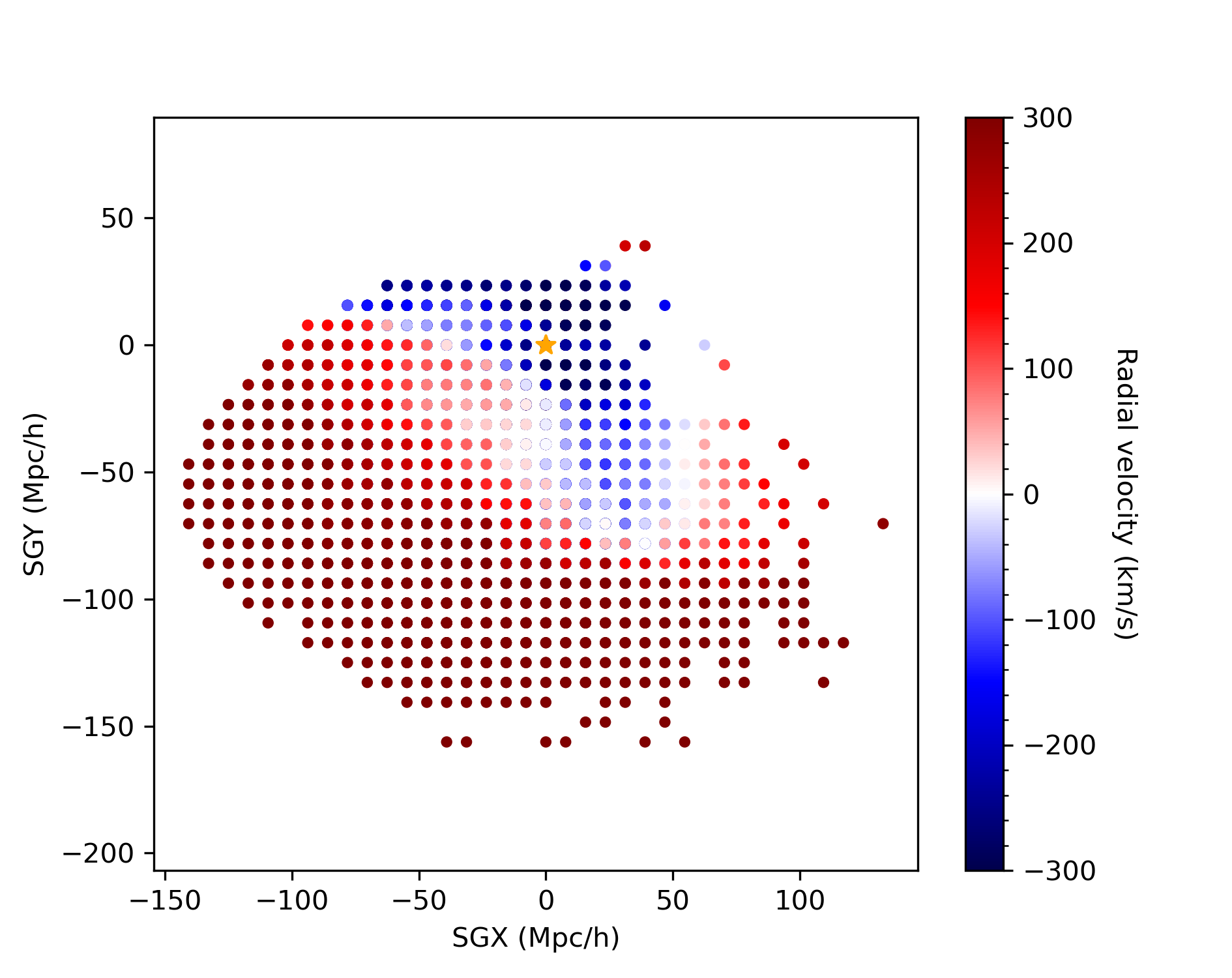}
	\caption{Radial peculiar velocity field of Laniakea after subtracting its Bulk Flow as seen by an observer located at the origin of the LCF, and its projection along orthogonal planes in the supergalactic Cartesian frame (an interactive version of this plot is available at \href{https://leolardo.github.io/Laniakea_Backreaction/}{https://leolardo.github.io/Laniakea\_Backreaction/}).}
	\label{RPV-Lan}
\end{figure}

To begin with, we compute the Bulk Flow (BF), the average velocity of the cells defining Laniakea, whose components turn out to be $v^x_{\rm{BF}}=-164 \pm 4$, $v^y_{\rm{BF}}= 19 \pm 2$, and $v^z_{\rm{BF}}=-96\pm 3$ km/s in the supergalactic Cartesian frame. Such a BF, in the picture of an expanding ellipsoid embedded in a spatially flat FLRW Universe, describes the relative motion of the inhomogeneity with respect to the cosmological rest frame (at the present time).  If we were to compute the radial velocity field of Laniakea with respect to its centre, this BF would manifest itself as a dipolar contribution. We subtract such dipole by subtracting the BF from every cell within Laniakea (note that the uncertainties on the BF value have a negligible impact on our final results). The reconstructed radial peculiar velocity field in the LCF $v_r^{\rm{rcst}}$ is plotted, for reference, in Fig.~\ref{RPV-Lan}.

\subsection{Effective model}\label{effectivemodel}
The notion of distance in General Relativity, and therefore in Cosmology, is observer-dependent due to the general covariance of the Einstein field equations.  In particular, the solution of the null geodesics equation for the metric in Eq.~\eqref{FLRWmetric} with $K=0$ gives the following definition of comoving distance $\bar{\chi}$  for a photon emitted at the time $t=t_e$ and observed today at $t=t_0$:~\footnote{In what follows we will refer to quantities computed in a background spatially FLRW metric through an overbar, $\bar{x}$.}
\begin{equation}\label{comovingdistanceflrw}
    \bar{\chi}=c\int_{t_0}^{t_e} dt \frac{1}{\bar{a}(t)} = c\int_0^{\bar{z}}  d\tilde{z} \frac{1}{H(\tilde{z})}\;,
\end{equation}
where in the second equality we have introduced the FLRW cosmological redshift $\bar{z}=\bar{a}(t_0)/\bar{a}(t_e) -1$, the FLRW Hubble parameter $\bar{H} = \dot{a}/a$ (where the dot denotes a derivative with respect to cosmic time), and adopted the common normalization choice $\bar{a}(t_0)=1$. Simply put, whilst the ``physical'' distance between two points in an expanding FLRW Universe evolves with time, their comoving distance as given by Eq.~\eqref{comovingdistanceflrw} remains constant. 

Most cosmological observations are based on two types of ``physical'' distances: the angular diameter distance $d_A$ and the luminosity distance $d_L$, which are properly defined in any space-time~\cite{Perlick:2010zh}. The former relates the physical length of an object to its subtended angular size as perceived by a distance observer, $d_A \equiv \delta A/\delta \Omega$, where $\delta A$ is the proper area of the emitting surface perpendicular to the line-of-sight, and $\delta \Omega$ the subtended angular element. The latter relates the bolometric (i.e.\ integrated along all frequencies) absolute luminosity  $L$ of the emitter with the bolometric flux $F$ measured by the observer. In any metric theory of gravity where photon number is conserved, the angular diameter and luminosity distances are related by Etherington's distance-duality relation, $d_L = (1+z_{\rm obs})^2 d_A$.~\footnote{We purposely used non-barred quantities here as the Etherington's reciprocity theorem, at the basis of the distance-duality relation, is true beyond the FLRW assumption, for the \textit{observed} redshift $z_{\rm {obs}} = (\lambda_{\rm{obs}} -\lambda_{\rm{emit}})/\lambda_{\rm{emit}}$, defined as the relative variation in the wavelengths of the photon from its emission to its observation.} In a FLRW Universe there is a very simple interpretation of the angular-diameter distance, $\bar{d}_A = a(t_e) \bar{\chi}$. In other words, the angular diameter distance is just the comoving distance we would observe today rescaled with the scale factor (hence the name) at the time light was emitted.
 
On the other hand, deviations from homogeneity and isotropy significantly complicate this picture. In particular, due to the peculiar motion of both the emitting source and the observer (induced by the different density fields at the respective positions) and the presence of inhomogeneities along the line-of-sight, the observed redshift $z_{\rm{obs}}\neq \bar{z}$ differs from the cosmological one induced by the expansion of the Universe only. Using linear order perturbation theory we can predict these effects from the knowledge of the density field, and therefore correct observed redshifts to recover the cosmological ones. Neglecting lensing and magnification effects (which are subdominant at lower redshifts, such as those relevant for Laniakea, compared to peculiar motions~\cite{2011ApJ...741...67D,Hui:2005nm}), we can write the following relation between the observed redshift $z_{\rm obs}$ and the cosmological one $\bar{z}$:
\begin{equation}
    \left(1+z_{\rm obs}\right)=\left(1+\bar{z}\right)\left(1+\frac{v_{\rm pec}}{c}\right)\left(1+\frac{v_{o}}{c}\right)\;,
\end{equation}
where $v_{\rm pec}$ and $v_{o}$ are the line-of-sight projections of the peculiar velocities of the source and the observer respectively, which are assumed to be small compared to the speed of light $v_{\rm pec}\ll c$.

The CF4 velocity reconstruction, and therefore Laniakea, has been defined collecting redshifts of galaxies whose luminosity distance from us is known through some cosmology-independent distance indicator. After correcting these redshifts $z_{\rm obs}$  for the observer peculiar motion (known for example from the CMB dipole), one then assumes a fiducial Cosmology (i.e. $\bar{E}(z)$, defined from $\bar{H}(z)\equiv H_0 \bar{E}(z)$) to compute their cosmological redshift $\bar{z}$ and peculiar velocity $v_{\rm pec}$. Combining the measured $v_{\rm pec}$'s, one then uses linear order perturbation theory to derive a reconstructed map of the peculiar velocity field in the nearby Universe.   

With the goal of describing effectively Laniakea's impact on observations, our main model building assumption is the existence of a line-of-sight dependent function $H_{\rm{los}}\left(\theta, \phi,z\right)$ such that:
\begin{equation}\label{modelbuilding}
 \frac{ c\left[\left(1+\bar{z}\right)\left(1+ \frac{v_{\rm{pec}}}{c}\right) - 1\right]}{\bar{d}_L} \equiv c\frac{z}{\bar{d}_L}\approx H_{\rm{los}}\left(\theta, \phi,z\right)\;,
\end{equation}
where $z$ is the observed redshift corrected for the observer peculiar motion, i.e. $(1+z)=(1+z_{\rm{obs}})/(1+v_o/c)$. 
In other words, we assume that all the cells in Laniakea are at rest (comoving) with respect to each other, and any relative velocity between them is induced by a different (anisotropic) expansion rate (rather than peculiar motions) $v\approx H_{\rm{los}} \;\bar{d}_L/(1+z) $.

Our triaxially expanding toy model describes the space-time region enclosed within the ellipsoid defined in Sec.~\ref{bestfitellipsoid} with a Bianchi I line element 
\begin{equation}
    ds^2= -dt^2 + a(t)^2dx^2 + b(t)^2dy^2 + c(t)^2dz^2\;,
\end{equation}
where $a,b,c$ are three different scale factors, whilst the exterior (which we also refer to, with abuse of language, as \textit{background}) is described by the usual flat FLRW \eqref{FLRWmetric} one. In a Bianchi I spacetime the cosmological redshift of light emitted by a source at the time $t=t_e$ along a line of sight parametrized by a versor $\vec{C}$ with components $\left(C_a,C_b,C_c\right)$  can also be written analytically as \cite{Schucker:2014wca}
\begin{equation}\label{redBI}
    1+z = \frac{\sqrt{\frac{C_a}{a^2(t_e)} + \frac{C_b}{b^2(t_e)} + \frac{C_c}{c^2(t_e)}}}{\sqrt{\frac{C_a}{a_0^2} + \frac{C_b}{b_0^2} + \frac{C_c}{c_0^2}}}\;,
\end{equation}
where the subscript $_0$ on the scale factors indicates their value at the time of observation.
We further assume that the anisotropies are small, and that the following relation between the Hubble rates associated to the scale factors inside and the background one holds
\begin{equation}
    H_i= \bar{H} + \Delta H_i \qquad\ \Delta H_i/\bar{H} \ll 1 \;, 
\end{equation}
where $i=a,b,c$.
Given the relatively low-redshift of the Laniakea, as mentioned earlier, we also neglect the redshift evolution of the anisotropies so that each $\Delta H_i$ is approximately constant.
Under these assumptions the functional dependence of the directional Hubble parameter $H_d$ is rather simple. Subtracting the average cosmological expansion $\bar{H}(\bar{z})$ from the fiducial cosmology, the triaxial expansion along the principal axes of the ellipsoid $(a,b,c)$ (with constant expansion rates $(\Delta H_a, \Delta H_b, \Delta H_c)$)  predicts that the velocity vector of a source $S$ with coordinates $(\tilde{x},\tilde{y},\tilde{z})$ in the LCF is given by:
\begin{equation}
    \vec{v}_{\rm{S}} = \left(v_{\tilde{x}},v_{\tilde{y}},v_{\tilde{z}}\right) = \left(\Delta H_a \tilde{x}, \Delta H_b \tilde{y}, \Delta H_c \tilde{z} \right)\;,
\end{equation}
resulting in the following radial velocity with respect to the centre of Laniakea
\begin{equation}\label{theovr}
    v_{\tilde{r}}^{\rm{th}} = \frac{\left(\Delta H_a \tilde{x}^2 +\Delta H_b\tilde{y}^2 +\Delta H_c\tilde{z}^2 \right)}{\sqrt{\tilde{x}^2 + \tilde{y}^2 +\tilde{z}^2}}\;.
\end{equation}
We want to find the values of $\Delta H_a,\Delta H_b,\Delta H_c$ which best describe the radial velocity field inside Laniakea, and evaluate the goodness of the fit. To do so, we run a Markov Chain Monte Carlo (MCMC) using the \textit{emcee}\footnote{\href{https://github.com/dfm/emcee}{https://github.com/dfm/emcee}} package \cite{2013PASP..125..306F} to maximize the Likelihood:\footnote{To determine the convergence of the MCMC exploration we follow the prescription given in \href{https://emcee.readthedocs.io/en/stable/user/autocorr/}{https://emcee.readthedocs.io/en/stable/user/autocorr/} and check the estimated autocorrelation time $\tau$ every $100$ steps for each chain, considering it convergent if the estimate has changed by less then $1\%$.}
\begin{equation}
\log{\mathcal{L}}=-\frac{1}{2}\sum_n\frac{\left(v^{n\;\rm{th}}_{\tilde{r}} -v^{n\;\rm{rcst}}_{\tilde{r}}\right)^2}{\sigma^2_n}\;,
\end{equation} 
where we have assigned flat priors $-9 < \Delta H_i < 9 \; \rm{km}/\rm{s}/\rm{Mpc}$  on the Hubble rates along the principal axes\footnote{We further verified that changing the prior to $\pm 10$ and $\pm 12$ does not significantly change the results.} and where the index $n$ runs along the $4079$ coordinates defining Laniakea.

The uncertainty associated to $v_r^{\rm{rcst}}$, as described in \cite{Dupuy:2023ffz}, consists of a linear dispersion from the reconstruction $\sigma^{v}_{\rm lin}$ obtained by averaging over multiple Hamiltonian Monte-Carlo (HMC) realizations, and a flat nonlinear component $\sigma_{\rm{nl}}$ accounting from the presence of non-linear structures that is the same for all the cells, which is one of the free parameters of the HMC exploration with best fit $\sigma_{\rm{nl}}=170\; \rm{km/s}$. Since our peculiar velocity field is smoothed on a grid of cells of volume $\sim \left(15.6 \;\rm{Mpc}\right)^3$, we smooth accordingly our theoretical prediction \eqref{theovr} averaging it over the 8 vertices of each cell. In table \ref{modelbestfits} we report the results of the analysis, in terms of best fits and reduced $\chi^2_\nu$, for three different choices of the dispersion: \textit{i)} the linear component only, \textit{ii)} the non-linear one only, \textit{iii)} the summation in quadrature of the two.    

\begin{figure}[]
    \centering
    \includegraphics[scale=0.6]{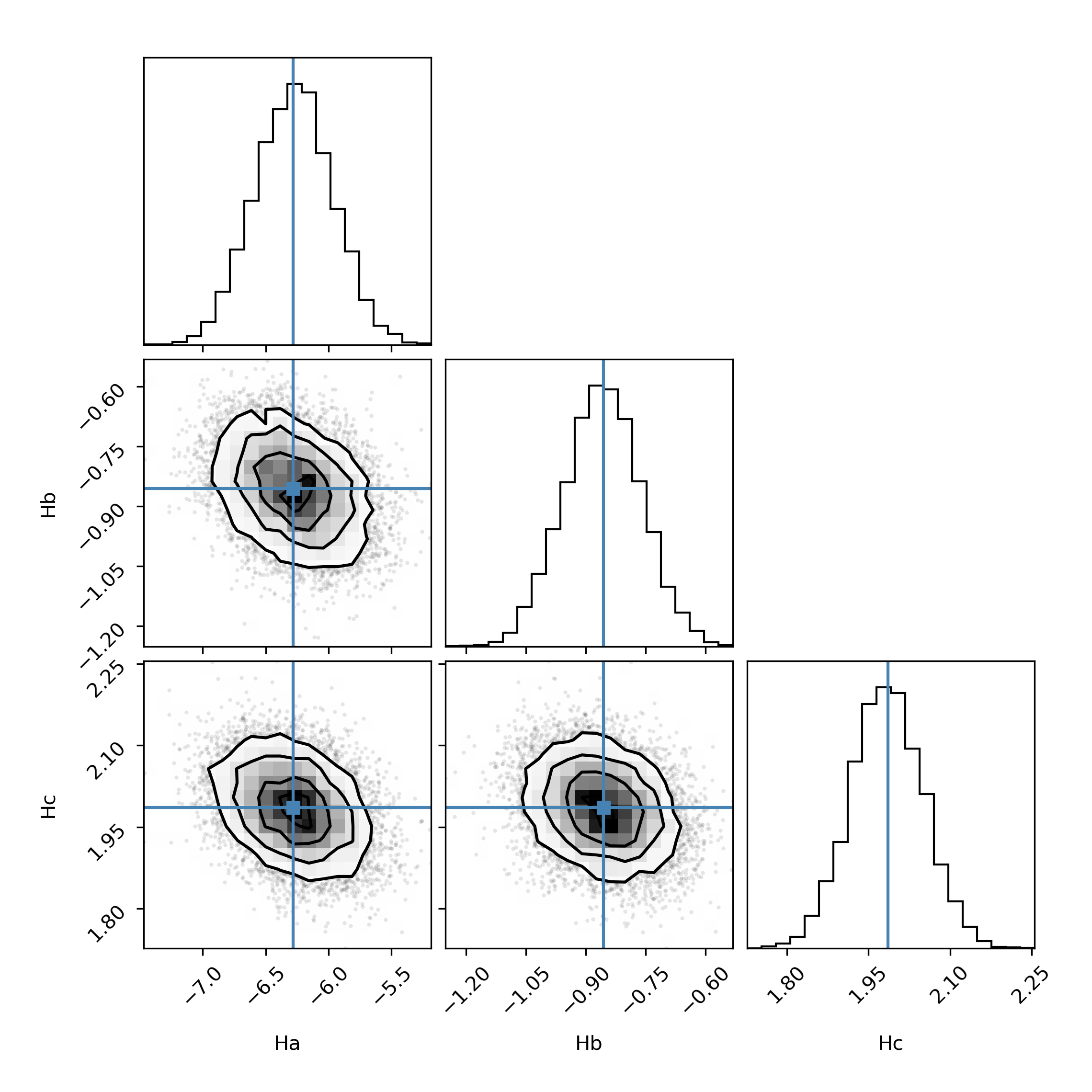}\\
	\caption{The Hubble rates along the principal semiaxes of the best fit ellipsoid from our MCMC analyses. The results displayed assume a constant velocity dispersion of $170$ km/s for each cell, corresponding to case \textit{ii)} from Table.~\ref{modelbestfits}}.  \label{Cornerplot}
\end{figure}

\begin{figure}[h]
    \centering
	\includegraphics[scale=0.6]{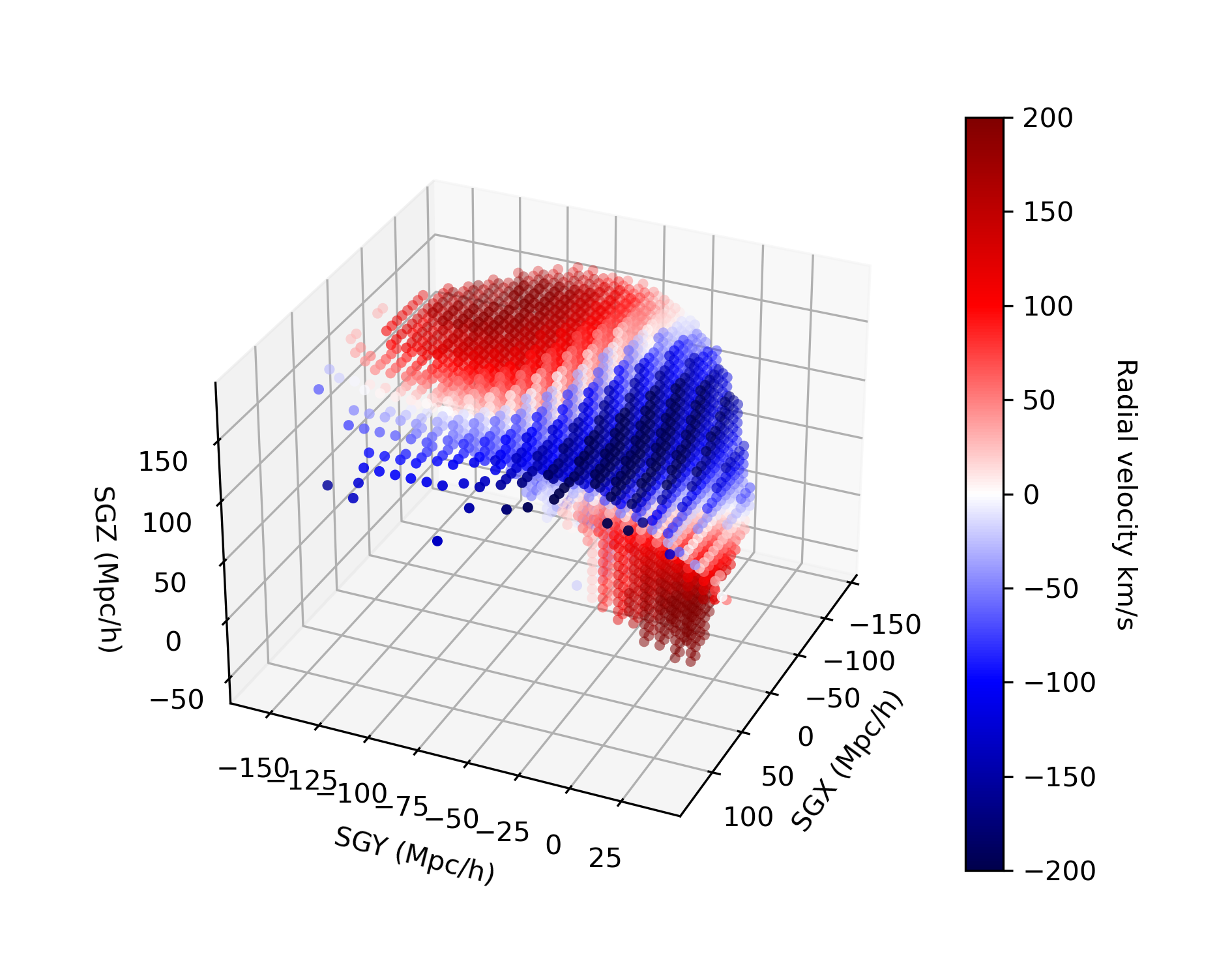}\includegraphics[scale=0.6]{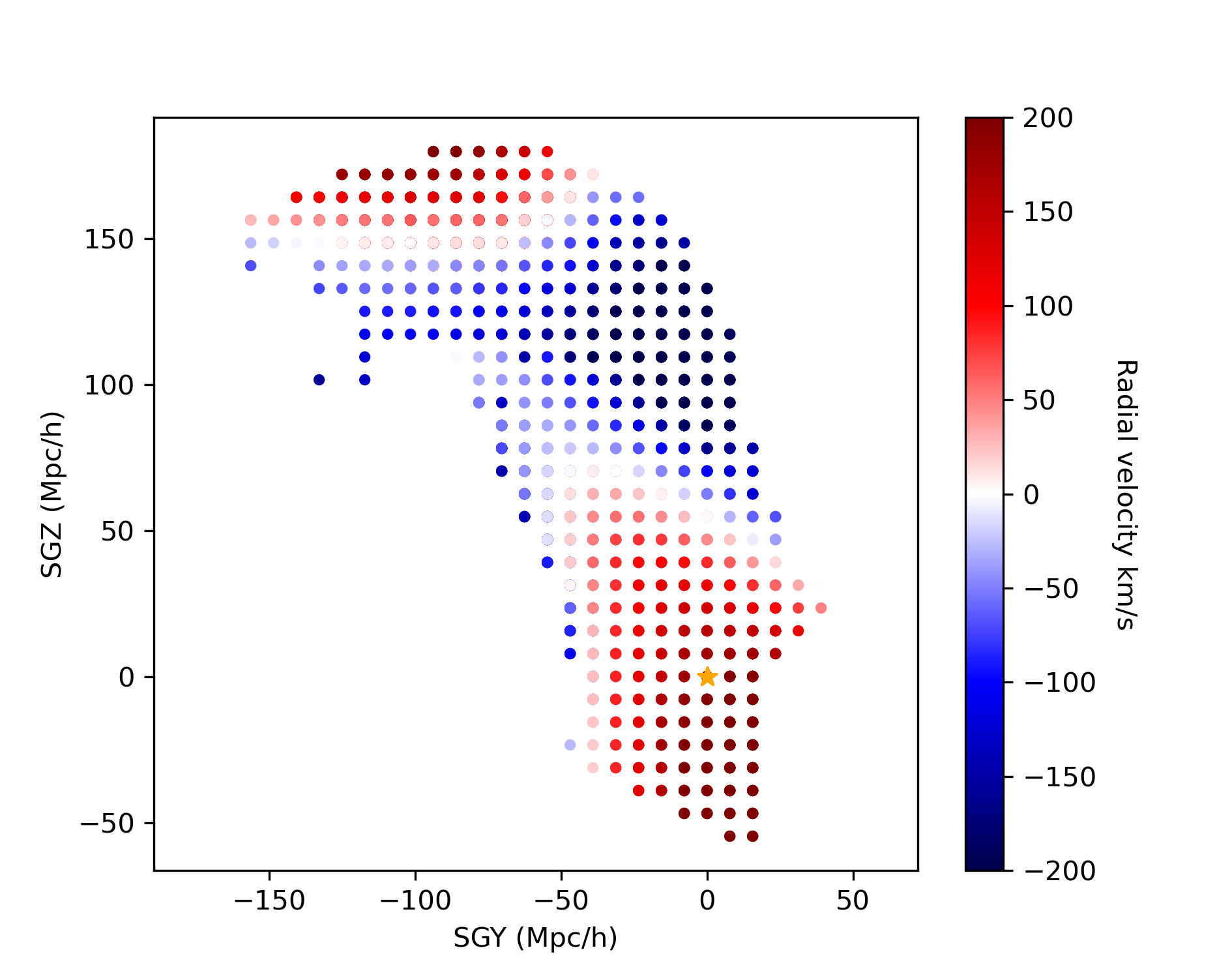} \includegraphics[scale=0.6]{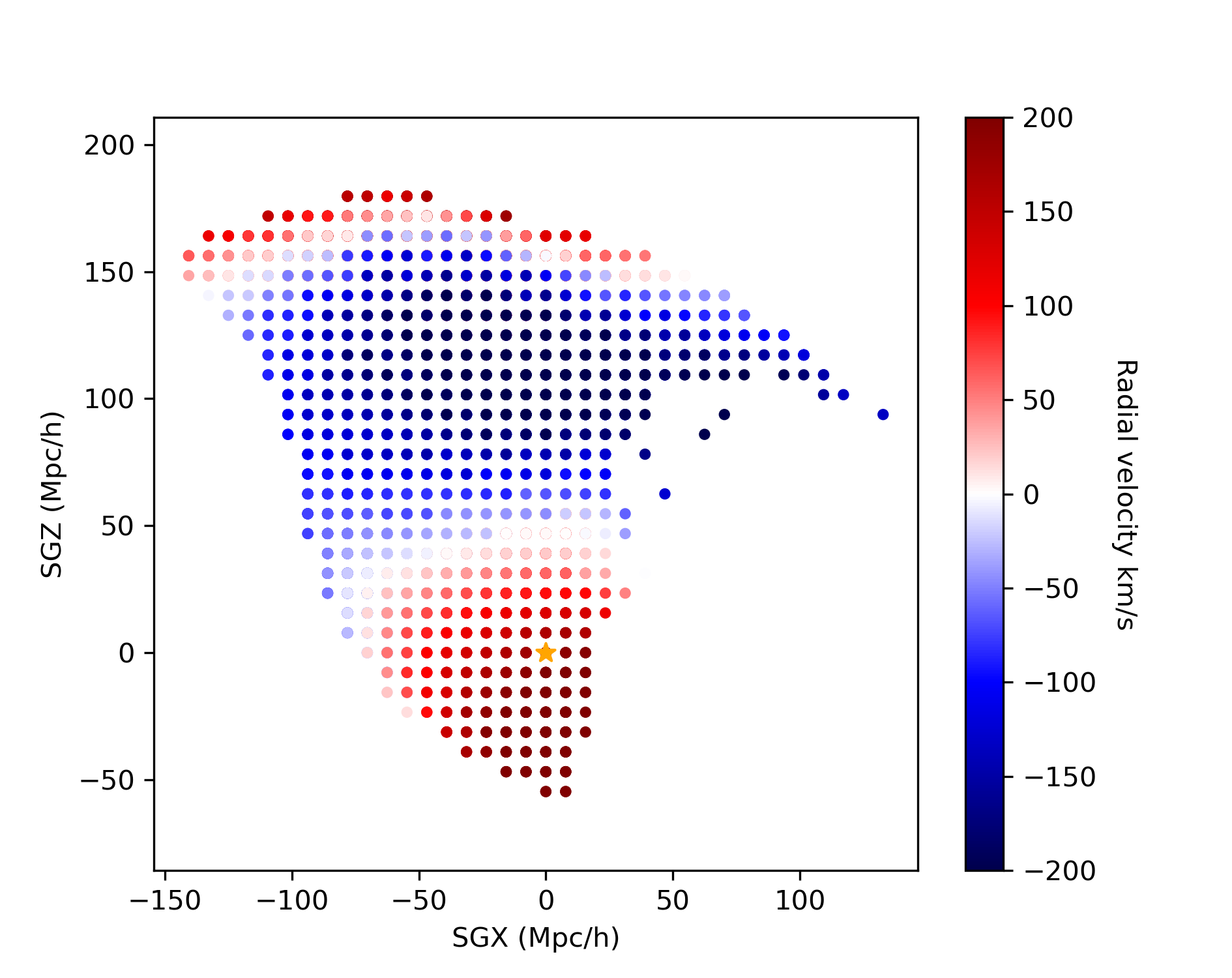}\includegraphics[scale=0.6]{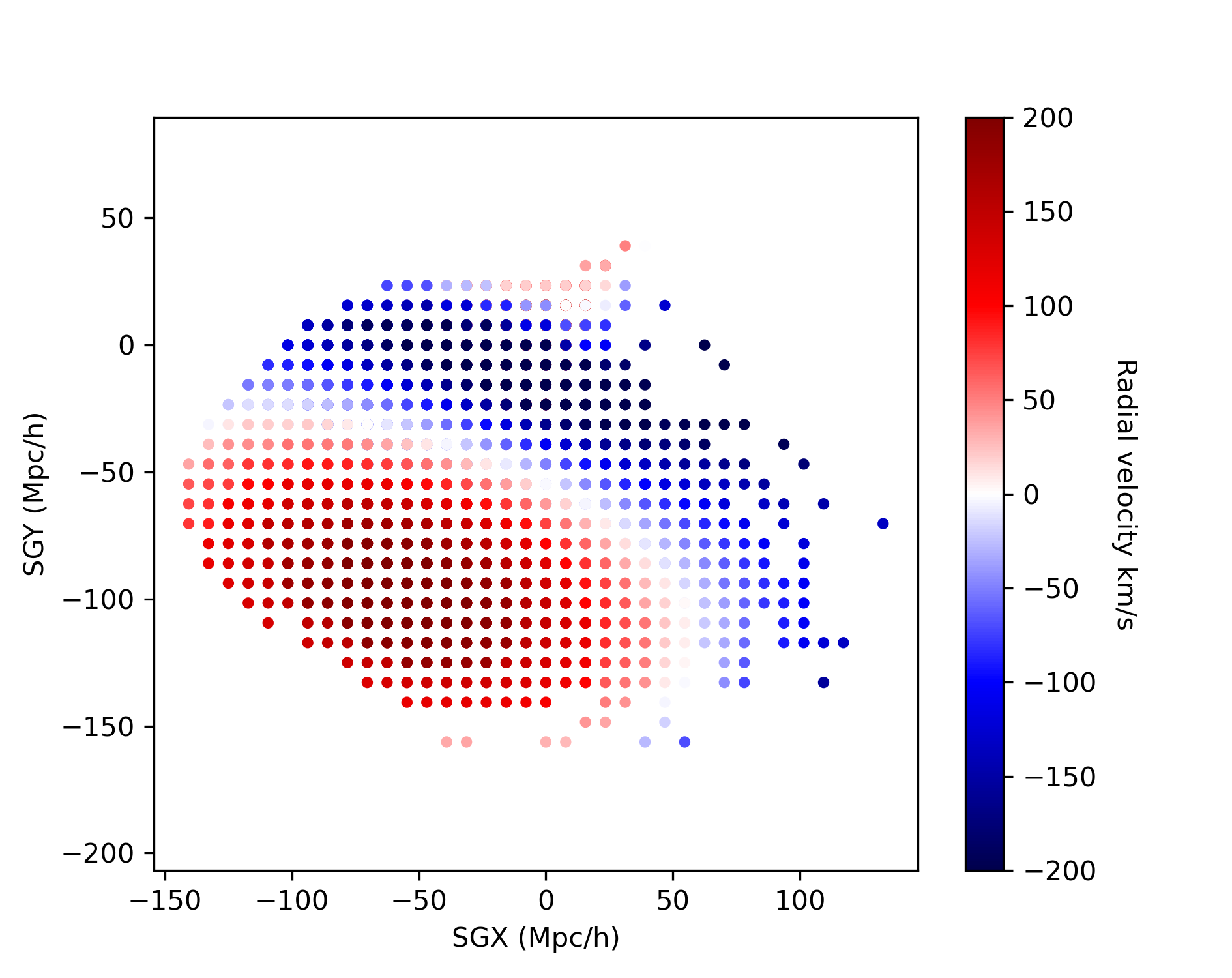}
	\caption{The theoretical prediction for the velocity field for the best fit values $\Delta H_a = -6.26$, $\Delta H_b = -0.86, \Delta H_c = 1.98 $, and its projection along orthogonal planes in the supergalactic Cartesian frame. (An interactive version of this plot is available  at \href{https://leolardo.github.io/Laniakea_Backreaction/}{https://leolardo.github.io/Laniakea\_Backreaction/}.)}
	\label{VelMap}
\end{figure}

\begin{table}[h]
\centering
\begin{tabular}{|c|ccccc|}
\hhline{|======|}
Dispersion & $\Delta H_a$ ($\rm{km}/\rm{s}/\rm{Mpc}$) & $\Delta H_b$ ($\rm{km}/\rm{s}/\rm{Mpc}$) & $\Delta H_c$ ($\rm{km}/\rm{s}/\rm{Mpc}$) & $\chi^2_\nu$ & p-value \\\hline
\textit{i)} $\sigma_{lin}$  & $-8.03 \pm 0.13$ & $-2.51 \pm 0.04$ & $2.18 \pm 0.02$ &  10.18 & $<10^{-6}$  \\
\textit{ii)} $\sigma_{nl}=170 \; \rm{km}/\rm{s}$  & $-6.14 \pm 0.26$ & $-0.48 \pm 0.08$ & $2.13 \pm 0.06$ & 1.41 & $\leq 10^{-6}$\\
\textit{iii)} $\sqrt{\sigma_{nl}^2 +\sigma_{\rm{lin}}^2}$  & $-6.26 \pm 0.31$ & $-0.86 \pm 0.10$ & $1.98 \pm 0.07$ & 1.07 & $3.9 \times 10^{-4}$ \\
\hhline{|======|}
\end{tabular}
\caption{The results of our MCMC parameter exploration for the triaxial rates of expansion $\Delta H_a,\Delta H_b,\Delta H_c$, and the corresponding chi-squared per degree of freedom  $\chi^2_\nu$ ($\nu =4076$)  for different dispersion modelling choices.}
\label{modelbestfits}
\end{table}
The reduced chi-square $\chi^2_{\nu}$ of the residuals can be used to evaluate the goodness of the fit. As expected, using only the linear component of the dispersion (case \textit{i)}) leads to a terrible fit, as it completely neglects the feedback of nonlinear structures on the first order peculiar velocities. Cases \textit{ ii), iii)} correspond to two extreme scenarios, as the former underestimates the error bars by neglecting the linear dispersion, and the latter overestimates them by neglecting the correlations between $\sigma_{\rm{lin}}$ and $\sigma_{\rm{nl}}$. One expects them to be anti-correlated, as one can always absorb a constant component of the linear dispersion into the non-linear one. Therefore the ``true'' uncertainty should lie between these two cases. On the other hand, case \textit{iii)} leads to the overall best fit (with a $\chi^2_{\nu}\sim 1.07$). For the rest of this work we will restrict ourselves to this dispersion modeling choice, for which the results of the MCMC exploration are reported in Fig.~\ref{Cornerplot}, and whose theoretical prediction for the velocity field is plotted in Fig.~\ref{VelMap}.

\begin{figure}[h]
    \centering
	\includegraphics[scale=0.6]{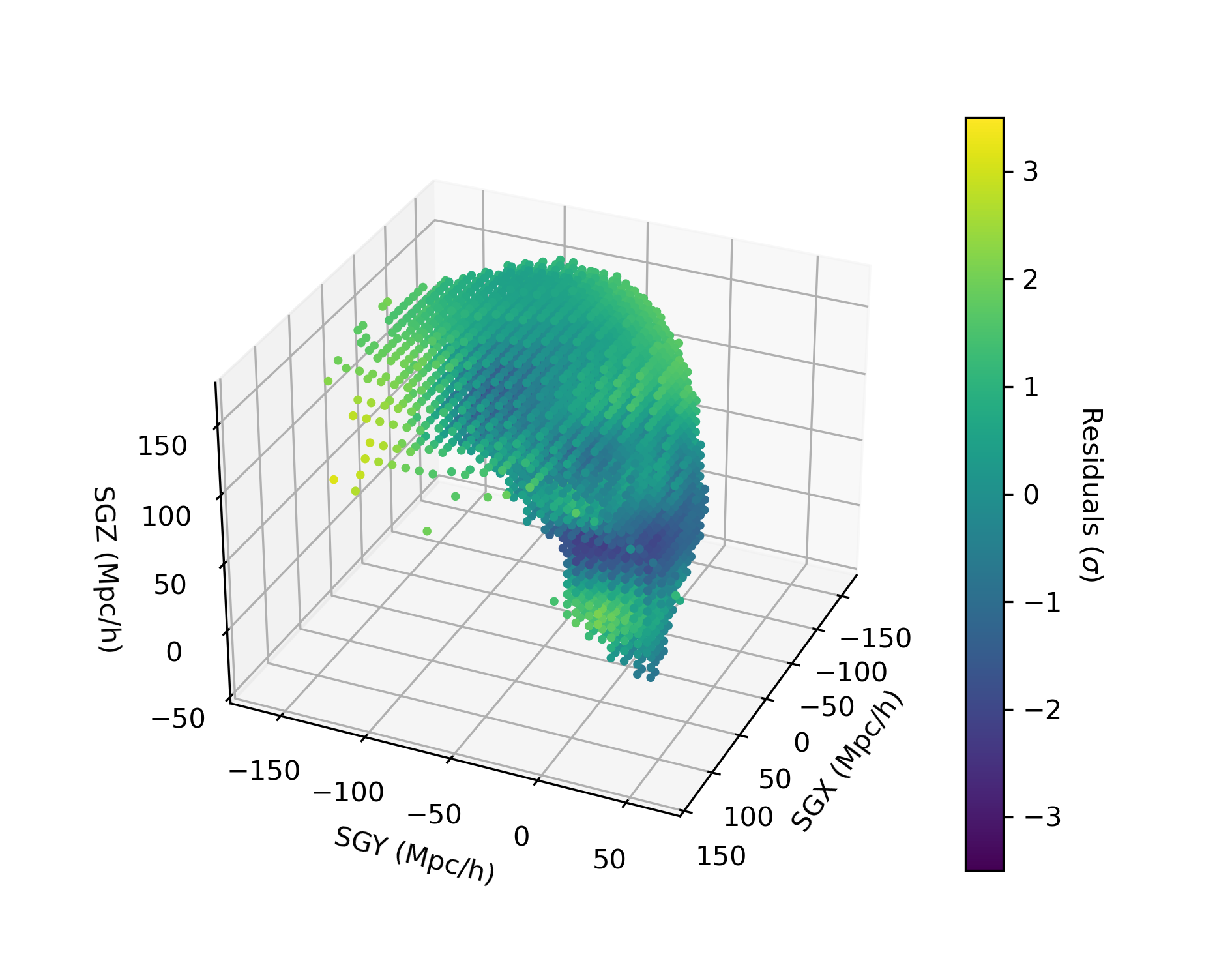}\includegraphics[scale=0.6]{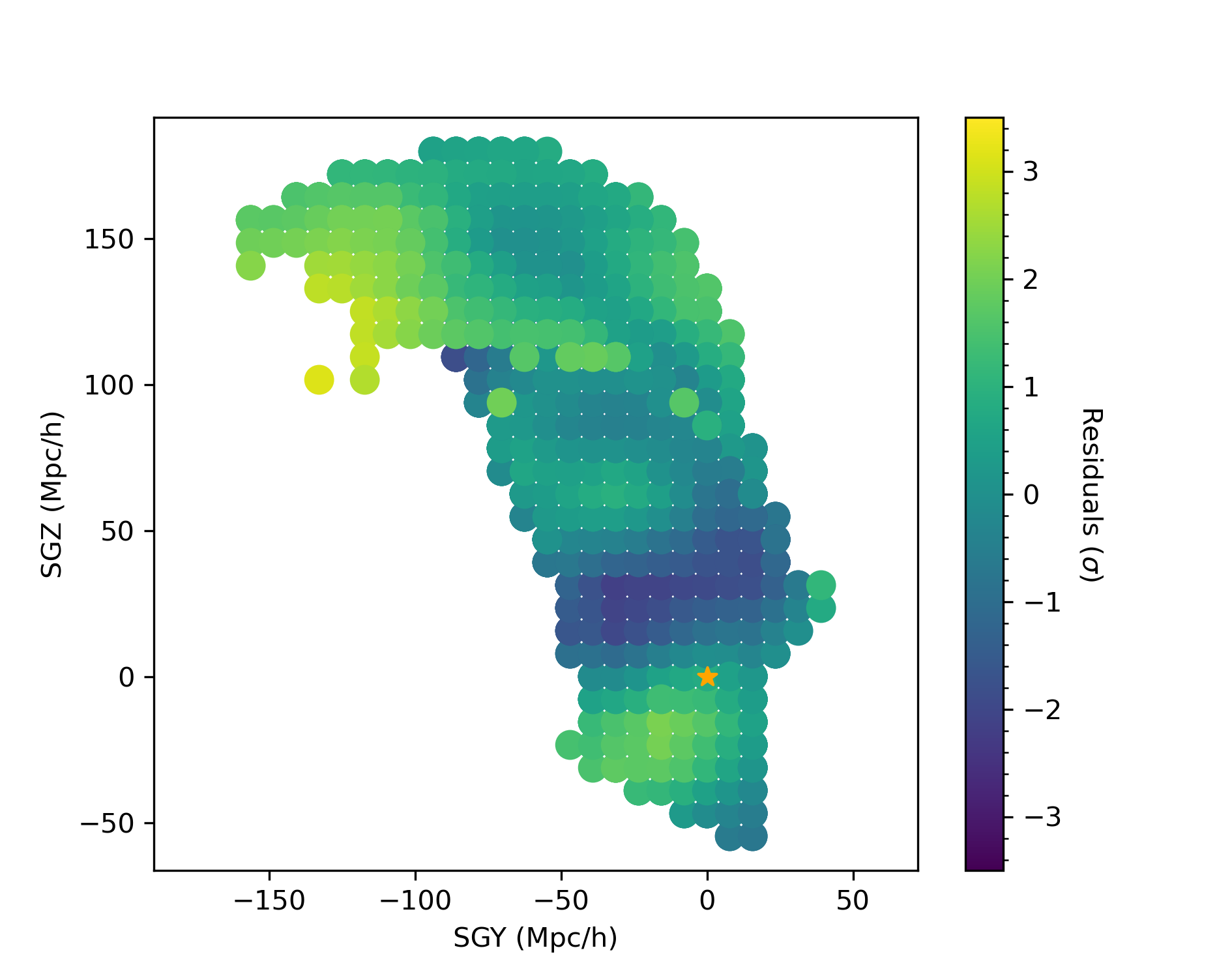}\\
 \includegraphics[scale=0.6]{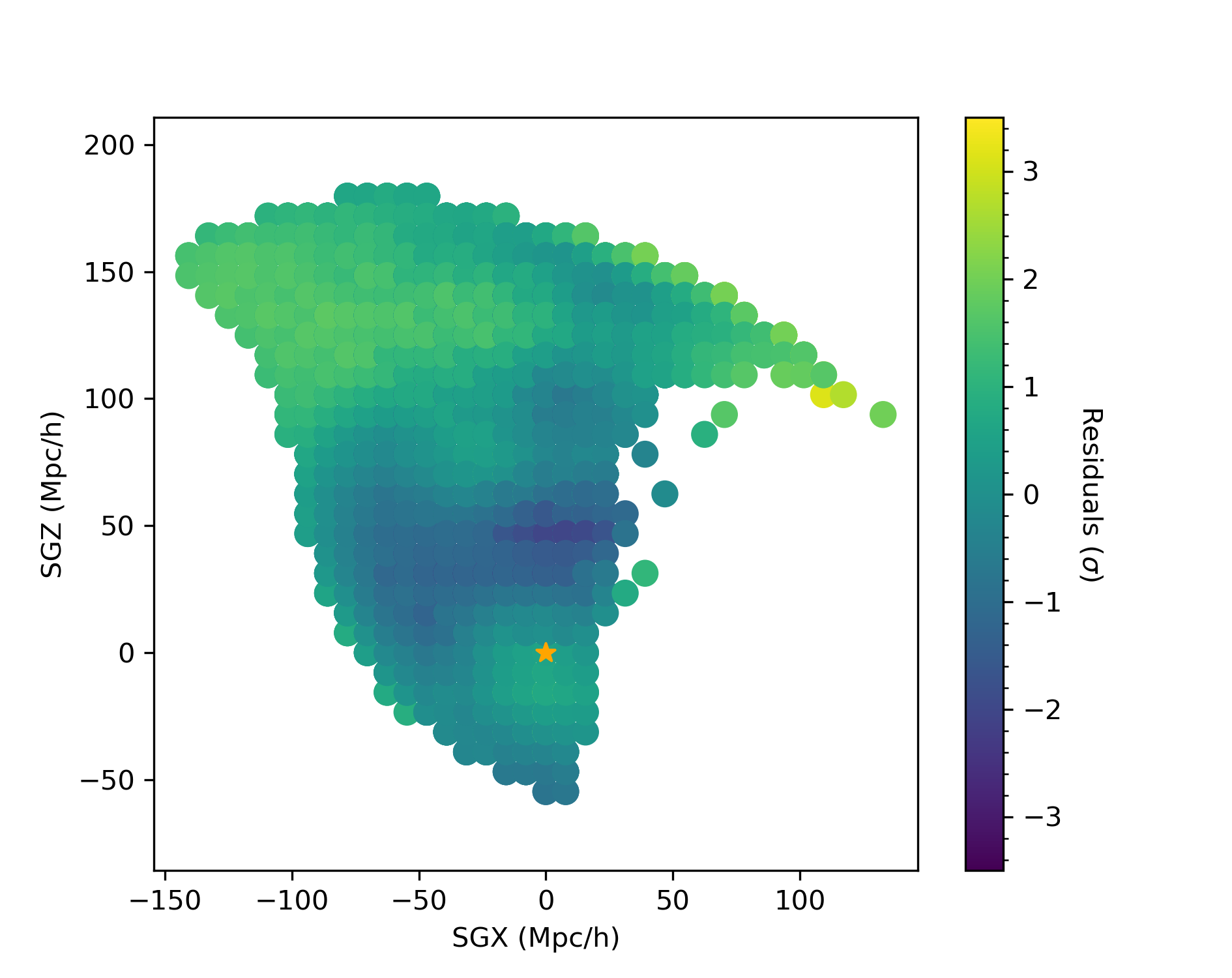}\includegraphics[scale=0.6]{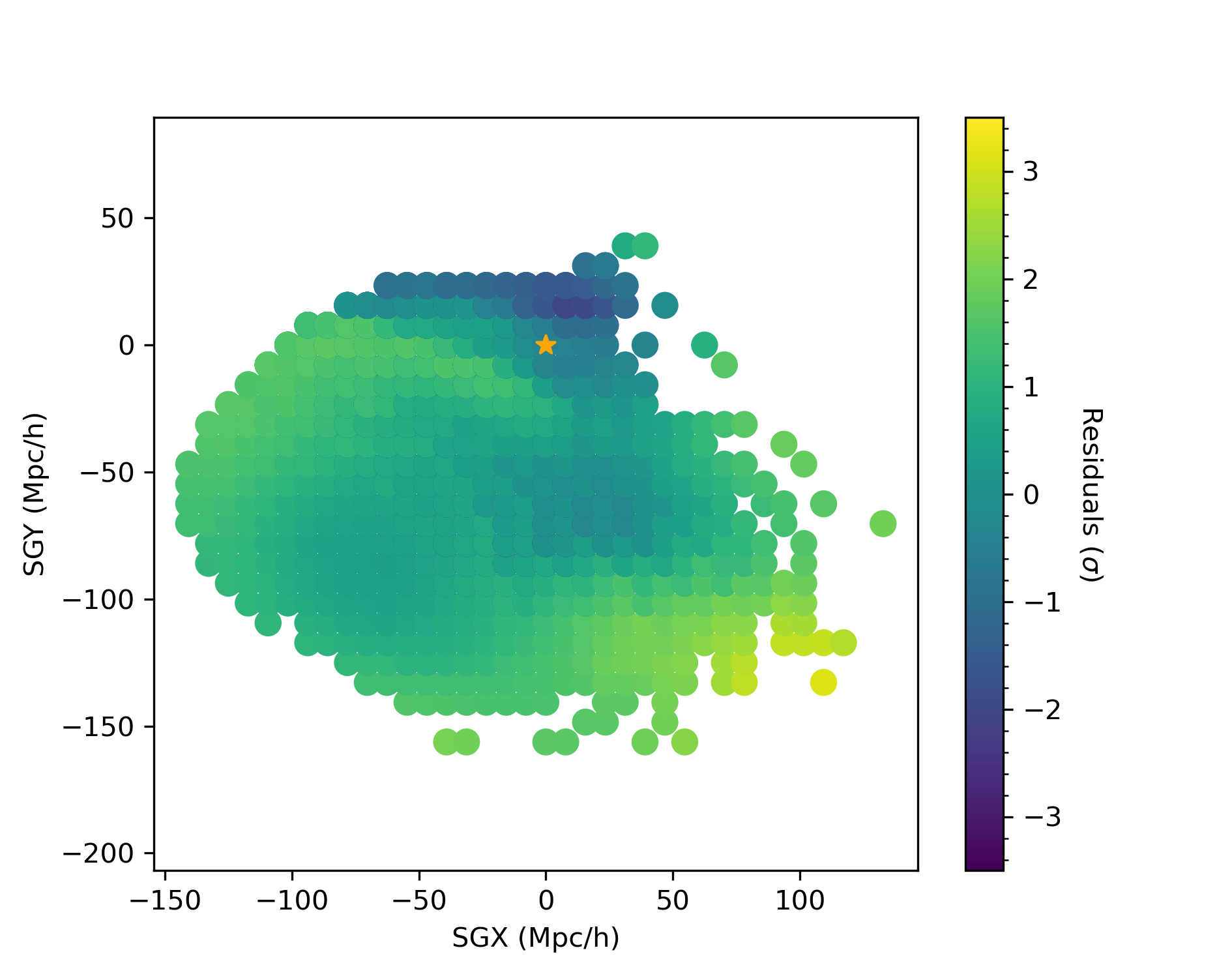}
	\caption{The spatial distribution of the residuals $v^{n\;\rm{th}}_{\tilde{r}} -v^{n\;\rm{rcst}}$ in units of standard deviations. Their spatial distribution, despite being compatible with a Gaussian (see Fig.~\ref{ResCen}), is non-random and most likely due to the presence of internal structures which are not captured by our simple model. The Milky Way is marked by an orange star. (An interactive version of this plot is available at \href{https://leolardo.github.io/Laniakea_Backreaction/}{https://leolardo.github.io/Laniakea\_Backreaction/}.)}
	\label{ResDistSpace}
\end{figure}

\begin{figure}[h]
    \centering
	\includegraphics[scale=0.8]{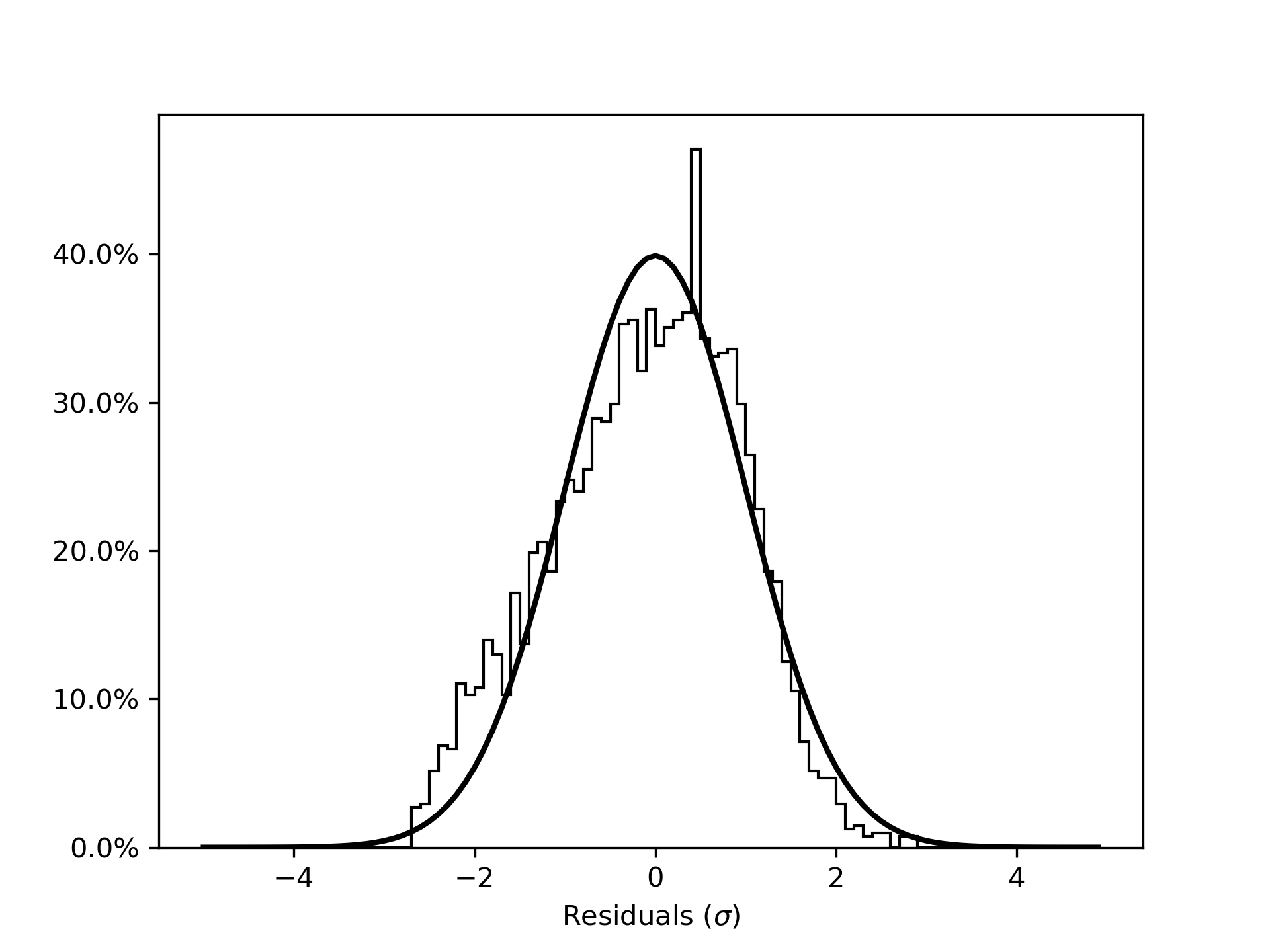}\\
	\caption{A normalized histogram of the residuals in units of standard deviations for the interior of Laniakea fitted by a triaxially expanding ellipsoid. For easier comparison with a normal distribution we plot on top of the histogram a normalized Gaussian with standard deviation $\sigma=1$.}
	\label{ResCen}
\end{figure}

In Fig.~\ref{ResDistSpace} we plot a 3-dimensional map of the spatial distribution of the residuals, and in Fig.~\ref{ResCen} their histogram in units of standard deviations. Whilst the histogram suggests that the scattering of the residuals is Gaussian, we see from their spatial distribution that they are non-random. A plausible explanation for such distribution is the presence of internal structures not captured by the homogeneous density profile assumed for the ellipsoid, suggesting that higher order multipoles might be needed to fully capture its behaviour. On the other hand, the fit remains excellent due to the large typical uncertainties of the peculiar velocities, and from now on we will assume it to be a good effective description of the internal dynamics of Laniakea.   


\subsection{Testing the homogeneity inside Laniakea}
\begin{figure}[h]
    \centering
\includegraphics[scale=0.8]{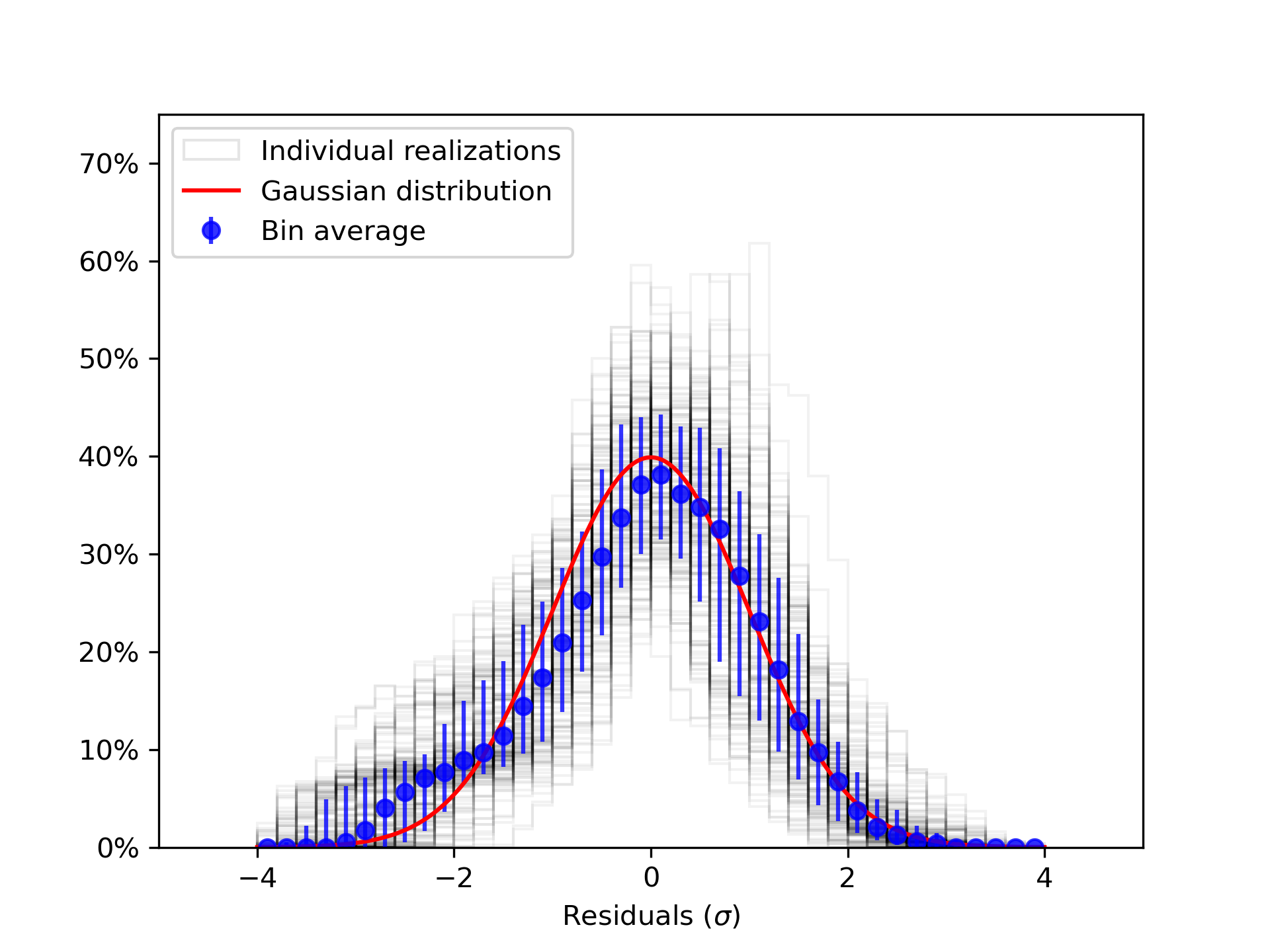}\\
	\caption{The distribution of the residuals between the reconstructed and predicted radial velocities with respect to 200 randomly chosen cells within Laniakea. Each individual distribution is plotted in black, and their binned average represented by the blue dots, with error bars indicating the $16^{\rm{th}}$ and $84^{\rm{th}}$ quantiles within the bin. We also plot for easier comparison a Gaussian distribution with unit variance (solid red line).  }
	\label{Homtest}
\end{figure}

Just as in a FLRW Universe, the assumed spatial homogeneity inside Laniakea implies that the relative velocities between any two pairs of points depend on their relative distance and orientation, but not from their particular spatial location. In other words, if we consider two sources and translate them along the same line-of-sight whilst keeping their distance fixed, their relative velocity should not change. To test the validity of this working assumption, we randomly pick 200 cells within Laniakea and for each of them compute the residuals between the predicted and reconstructed peculiar velocity of any other cell in the map using the best fit values of Table~\ref{modelbestfits} (case \textit{ii}). We plot in Fig.~\ref{Homtest} the histogram of their distribution for each realization (black contours), together with their bin average and the $16^{\rm{th}}$ and $84^{\rm{th}}$ quantiles (blue dots and corresponding error bars). Comparing the latter with a Gaussian (solid red line) we see that their distribution is skewed towards the right, as expected due to the non linear motion sourced by smaller substructures within the volume of Laniakea.  On the other hand, deviations from Gaussianity are relatively small, and we conclude (\textit{a posteriori}) that the working assumption of homogeneity is a reasonable one. 

\subsection{Comparison with a spherical model}
 It is interesting to compare our results to those we would obtain averaging over the anisotropies, i.e. assuming an homogeneously expanding spherical model $\mathcal{S}$ centered at the origin of the LCF. Running a similar MCMC analysis for a single parameter $\Delta H_{\rm{sph}}$ we find:
 \begin{equation}
     \Delta H_{\rm{sph}} = 0.30 \pm 0.04 \;\;\qquad  \chi^2_{\nu\;\rm{sph}}= 1.32\;,
 \end{equation}
and comparing the Bayesian Information Criterion (BIC) for the ellipsoidal and spherical models we find strong evidence that the triaxial model for Laniakea is preferred.\footnote{It is important to stress that allowing for an inhomogeneously expanding model \textit{a la\'} LTB might change the result. However, a comparison with such a model goes beyond the scope of the present work.}  
Interestingly, the best fit for $\Delta H_{\rm{sph}}$ is positive, in contrast with the fact that the average expansion rate within the ellipsoidal model is negative. This simple exercise highlights the relevance that local anisotropies might have on parameter estimation of LSS properties.


\section{Weak Backreaction of Laniakea on Cosmological observables}

\label{Backreaction_cosmo}

\subsection{Comoving and Luminosity distances}

\label{distances_comparison}
Now that we have developed a simple but robust effective model to describe the interior of Laniakea, we are in a position to estimate its impact on cosmological observables and, in particular, distances. For the remainder of this section, an upper bar on any cosmological quantity $\bar{x}$ refers to it as given in a flat FLRW Universe, whereas non-barred cosmological quantities $x$ are computed in our toy model for Laniakea. We are interested in comparing the inferred distance to the same object in the two models, where with ``same object'' we mean that the observed ``cosmological'' redshift of the source is the same $\bar{z}=z$.\footnote{In appendix \ref{appA} we perform a similar comparison but identify the ``same'' object as the one which emitted the photon at the cosmological time $t=t_e$, and show that the two results are consistent if one allows $t_e$ to be different but sets $z=\bar{z}$.}   

In a FRLW Universe the luminosity distance $\bar{d}_S$ to a source $S$ is given by:
\begin{equation}\label{dflrw}
    \bar{d}_S\left(z_S\right)=\left(1 + z_S\right)\bar{\chi}_S(z_S)\;,
\end{equation}
where $\chi$ is the comoving distance defined in Eq. \eqref{comovingdistanceflrw}.
Let us consider two points $A$ and $B$ on the same line-of-sight, with $B$ further away than $A$. It is convenient to rewrite the FLRW luminosity distance to $B$ as:
\begin{equation}
    \bar{d}_B(z_B)= \left(1+z_B\right)\left(\bar{\chi}_A(z_A) +\bar{\chi}_{AB}(z_A, z_B) \right)\;,
\end{equation}
where the quantity $\bar{\chi}_{AB}\left(z_A,z_B\right)$ is defined as
\begin{equation}
    \bar{\chi}_{AB}\left(z_A,z_B\right)= c\int_{z_A}^{z_B}d\tilde{z} \frac{1}{\bar{H}(\tilde{z})}\;,
\end{equation}
(these are a trivial re-writing of Eqs.~\ref{comovingdistanceflrw} and~\ref{dflrw}).

Of course the computation of these distances changes in our toy model, which predicts a direction-dependent expansion rate $H(z,\theta,\phi)$ inside Laniakea. On the other hand, since we are working with small anisotropies restricted to the interior of Laniakea, let us assume that if the point $A$ lies on its boundary then the luminosity distance to $B$ can be written as the sum of two terms
\begin{equation}\label{dlan}
\begin{split}
    d_B\left(z_B,\theta,\phi\right)&= \left(1+z_B\right)\left(c\int_{0}^{z_A}d\tilde{z}\frac{1}{H\left(\tilde{z},\theta,\phi\right) }+c\int_{z_A}^{z_B}d\tilde{z}\frac{1}{\bar{H}\left(\tilde{z}\right) }\right)\; \\
    &=\left(1+z_B\right)\left[\chi_A(z_A,\theta,\phi) + \bar{\chi}_{AB}\left(z_A,z_B\right) \;\right],
\end{split}
\end{equation}
where we have introduced a line-of-sight dependent ``generalized'' version of the FLRW comoving distance:\footnote{Note that with the definition of ``comoving'' distance of Eq.\eqref{BIdist}, as long as we can neglect the redshift evolution of the anisotropies (as we are assuming in our toy model), the luminosity distance 
 $d_L \equiv (1+z)\chi$ is consistent with the one for a Bianchi I Universe derived in Ref.~\cite{Schucker:2014wca}, Eqs.~(108) and (111), in the limit of small redshifts and eccentricities.}
\begin{equation}\label{BIdist}
    \chi_A(z_A,\theta,\phi) \equiv c\int_{0}^{z_A}d\tilde{z}\frac{1}{H\left(\tilde{z},\theta,\phi\right) }\;.
\end{equation}
It is important to stress that the above effective description, in particular the decomposition of the integral in Eq. \eqref{dlan}, neglects the impact of the discontinuity of the metric at the boundary of Laniakea in our toy model. On the one hand, this is unavoidable as it is not fully understood how to embed an anisotropic Bianchi inhomogeneity into a FLRW background apart from very particular cases. In a realistic scenario, one should expect some smooth continuous transition to occur around the boundary between the two metrics, which can be mimicked by approximating the step functions between the two region with a boundary of sharp hyperbolic tangents. On the other hand, given the practical nature of the effective description proposed here, we will assume that these boundary effects are small compared to the first order effects described here.

Let us now focus our attention to the following crucial question: for sources $B$ to which we assign a cosmological redshift $z_B$ (for example after correcting for their and our own peculiar motions), how different would be our inference of their luminosity distances in a Universe with or without Laniakea?
To answer this question within our toy model, let us consider the difference between Eqs.~\eqref{dlan} and \eqref{dflrw}. It is useful to define the quantity $\Delta \chi_A$
\begin{equation}\label{deltachi}
    \Delta \chi_A(\bar{z}_A,\theta,\phi) = c\int_0^{\bar{z}_A}d\tilde{z}\frac{1}{H\left(\tilde{z},\theta,\phi\right)} -\frac{1}{\bar{H}(\tilde{z})}\;,
\end{equation}
which represents the change in the comoving distance to a point $A$ at the boundary of Laniakea obtained using the expansion rate in Laniakea rather than the FLRW one. In the previous section we found that for each line-of-sight $H_d(z,\theta,\phi)-\bar{H}(z)\equiv \Delta H(\theta,\phi)$ is at most of the order of a few km$/$s$/$Mpc, and therefore $\Delta H/\bar{H}$ can be treated as a perturbative quantity. After Taylor expanding around it and neglecting higher order terms we find 
\begin{equation}\label{d-d}
 d_B(z_B,\theta,\phi) - \bar{d}_B(z_B) \approx  \left(1+z_B\right)\Delta \chi_A(\bar{z}_A,\theta,\phi) \;,  
\end{equation}

from which we can finally rewrite the main contribution to the relative corrections of the luminosity distance in our effective model as
\begin{equation}\label{deltad}
    \frac{ d_B - \bar{d}_B}{\bar{d}_B}\equiv \frac{\delta d_B}{\bar{d}_B} \approx \frac{\Delta\chi_A}{\bar{\chi}_B}\;.
\end{equation}

\begin{figure}[]
    \centering
    \includegraphics[scale=0.8]{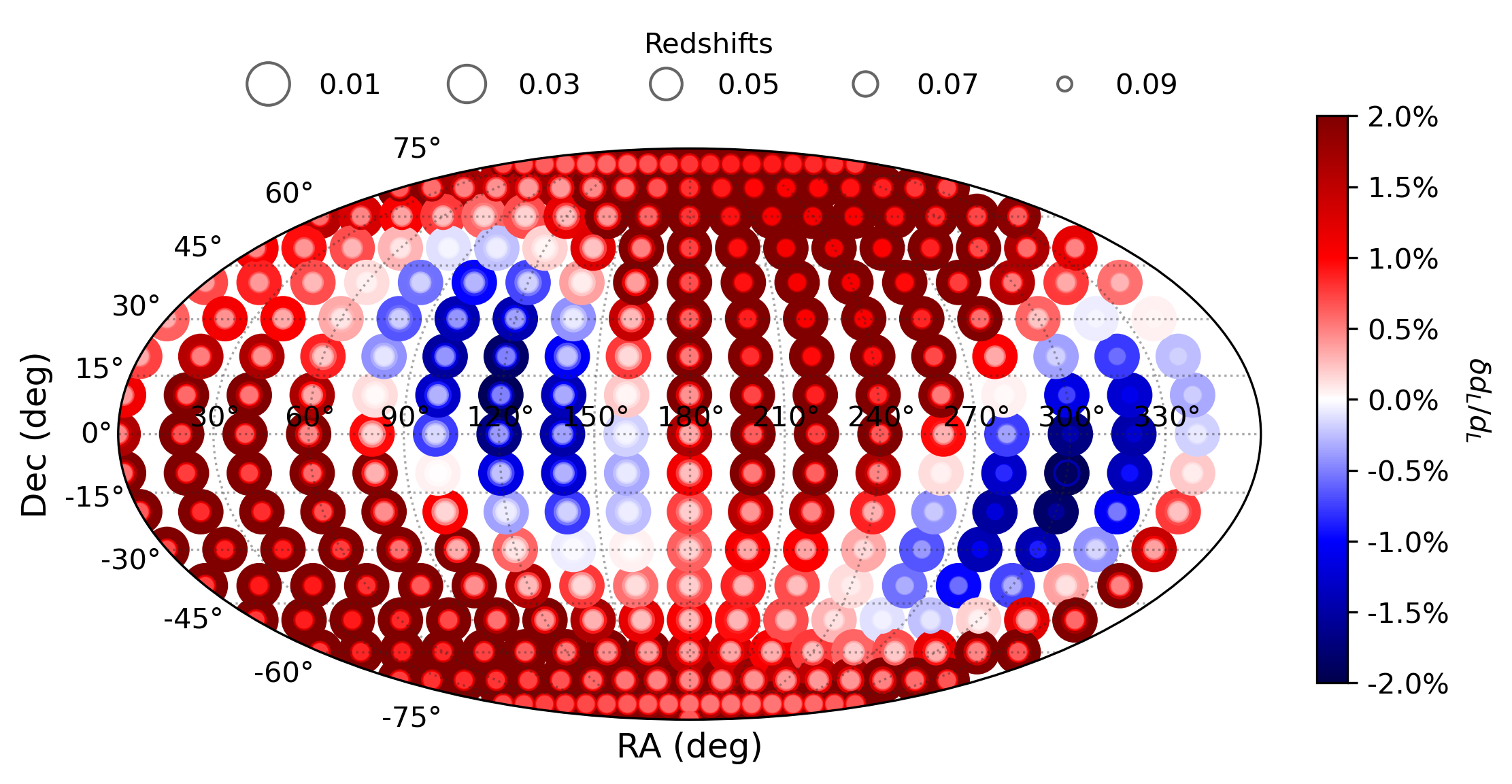}\\
	\caption{Relative corrections on the luminosity distance for a homogeneous and isotropic full-sky distribution of sources up to redshift $z=0.1$ in (RA,Dec). The size of the circles is inversely proportional to the redshifts of the sources, as indicate in the upper legend.}
 \label{corrfullsky}
\end{figure}

\begin{figure}[]
    \centering
    \includegraphics[scale=0.8]{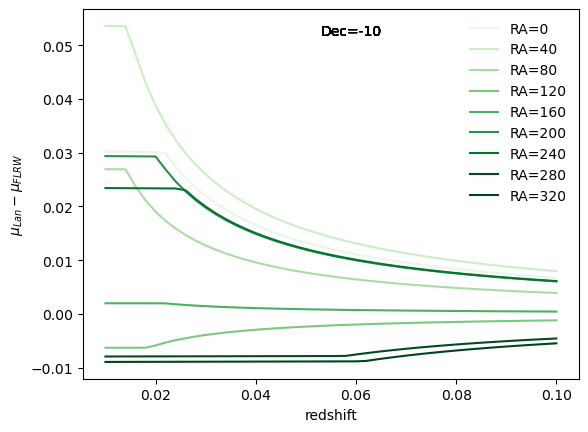}\\
	\caption{Corrections to the apparent magnitude of sources in the Hubble diagram line-of-sights with fixed Declination $\rm{DEC}=-10$ as a function of the redshift and the Radial ascention. The corrections are approximatively constant within the interior of the ellipsoid and vanish asymptotically outside.}
 \label{los-corrections}
\end{figure}

In Fig.~\ref{corrfullsky} we plot these corrections for a homogeneous and isotropic full-sky distribution of sources over the redshift range $0 < z < 0.1$, as seen by an observer in the Milky Way. We can identify two main concurring effects: the quadrupolar behaviour expected from a triaxial model with two contracting axes and an expanding one, and the hemispherical asymmetry induced by the observer's offset from the center of the ellipsoid. Because of the latter, the portion of sky closest to the edge of Laniakea will be less affected since photons will travel a shorter distance within the inhomogeneity (a similar effect has been suggested in Ref.~\cite{Perivolaropoulos:2023tdt} to explain an observed anisotropic distribution of low-$z$ sources in Pantheon+, but where the offset is from the center of a spherical distribution). A similar conclusion can be drawn from Fig.~\ref{los-corrections}, where the relative corrections are plotted as functions of redshifts for a number of representative line-of-sights.
For the remainder of this work, we will refer to those corrections for convenience as \textit{Laniakea's backreaction}.

\subsection{Low-$z$ SN~Ia and impact on $H_0$}
To assess the impact of Laniakea's backreaction on the determination of $H_0$ from low-$z$ SN~Ia we compare redshifts and distance moduli measurements from the Pantheon+ catalog \cite{Scolnic:2021amr} with our prediction for the distances Eq.~\eqref{deltad}.

\begin{figure}[h]
    \centering
	\includegraphics[scale=0.8]{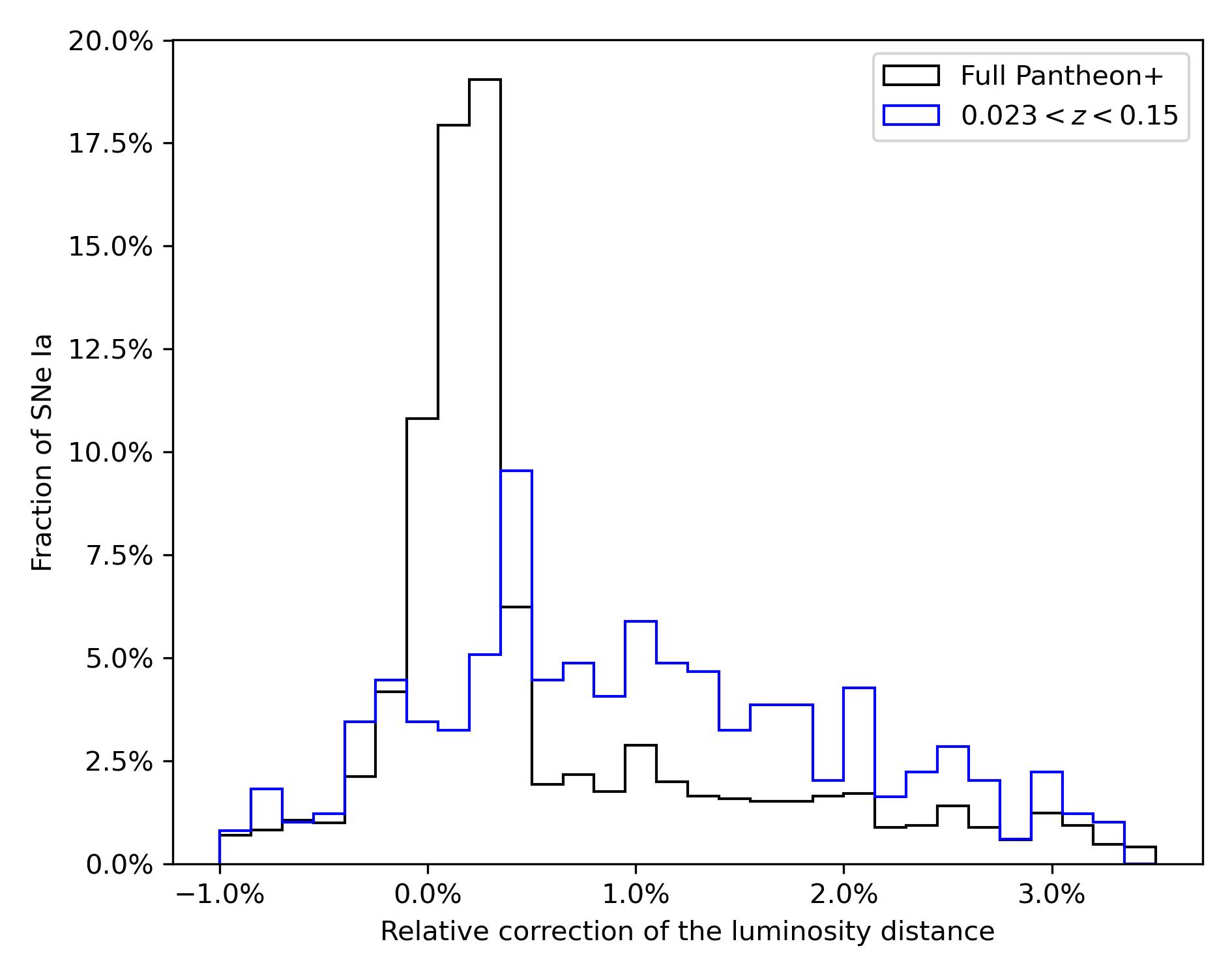}\\
	\caption{The distribution of the relative corrections of the luminosity distance induced by Laniakea's backreaction on SN~Ia from the Pantheon+ catalog. The black line represents the total Pantheon+ sample, whilst the blue line only those used by the SH0ES collaboration for the determination on $H_0$.}\label{Pantheon+_corr}
\end{figure}

\begin{figure}[]
    \centering
    \includegraphics[scale=0.78]{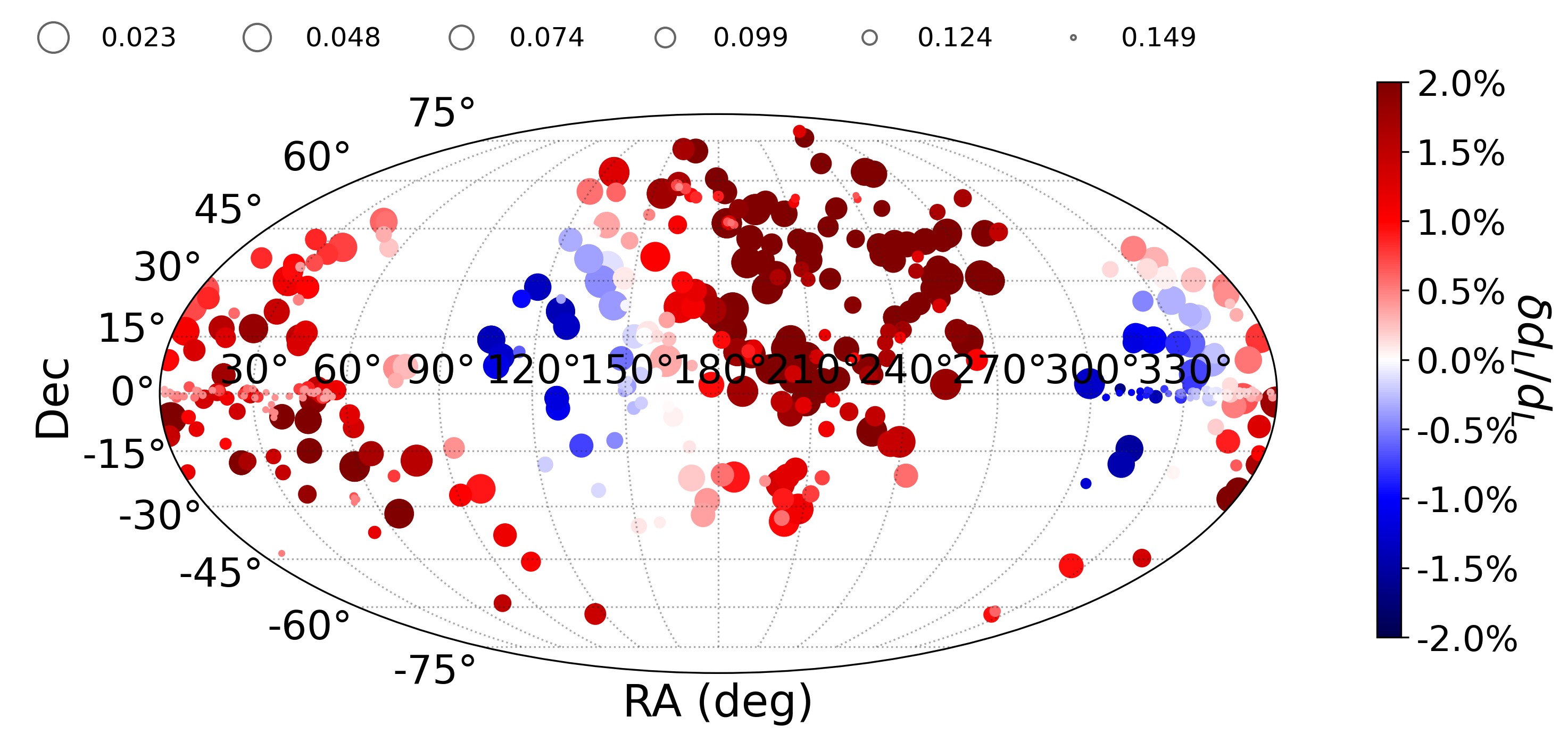}\\
	\caption{The position in the sky of the low-$z$ subset of SN~Ia in Pantheon+, color-coded with the amplitude of the relative correction to the luminosity distance induced  by Laniakea's backreaction.}\label{P+lowzcorr-radec}
\end{figure}

Since our effective model has been obtained through the analysis of peculiar velocities, to avoid bias induced by double counting of corrections, we use redshifts corrected for the kinematic CMB dipole but not for PV. In Fig.~\ref{Pantheon+_corr} we plot the histogram of the relative corrections of the luminosity distance of each supernova in the sample and for those in the redshift range $0.023 < z < 0.15$, some of which are used by the SH0ES collaboration in Ref.~\cite{Riess:2021jrx} to measure $H_0$. As expected, since the relative corrections are inversely proportional to the distances from the sources, their impact is completely negligible ($<0.1\% $) for most of the supernovae in the full sample. On the other hand, the histogram for the low-$z$ subset exhibits a moderate skewness towards higher values of the relative correction, which can be as large as $3\%$.
In Fig.~\ref{P+lowzcorr-radec} we plot the corrections on the luminosity distance of  the low-$z$ supernovae as function of their position in the sky in RA and DEC.

To compute the impact that these corrections have on the $H_0$ determination, we used the distance modulus $\mu$ from the low-$z$ subset of Pantheon+ SN~Ia and adopted, for simplicity, only the diagonal part of the covariance matrix for the errors.\footnote{The main reason behind this choice is that the full covariance matrix already incorporates corrections induced by peculiar velocities. Our toy model predicts that inside Laniakea all the sources are at rest (comoving), so that we should get rid of such corrections (and their correlations with other sources of uncertainty) for objects in its interior. Consistency would then require to recompute the full Pantheon+ covariance matrix within these hypotheses, which goes beyond the scope of the present work.} $H_0$ is then obtained by minimizing the sum of the residuals between the luminosity distances derived from the observed distance modulus and the low-$z$ cosmographic expansion of $d_L$ \cite{Visser:2004bf} up to quadratic order, assuming a deceleration parameter $q_0 = -0.55$.  More specifically, we assume that in our toy model the distance modulus of a source $i$ can be written
\begin{equation}
    \mu_i = 5\log{\left\{\frac{cz_i}{H_0}\left[1+\frac{1}{2}\left(1-q_0\right)z_i\right] \left(1+\frac{\Delta \chi_i}{\chi_i}\right)\right\}    }+25\;,
\end{equation}
where $\Delta \chi_i$ represents the change in the comoving distance from the observer to the boundary of Laniakea along the line-of-sight to the source $i$, as in Eq.~\eqref{deltad}.
 We found that the impact on the best fits for $H_0$ obtained with and without corrections is an effective shift of order $\Delta H_0 = H_0^{\rm{Lan}} -\bar{H}_0 \simeq 0.6~\rm{km}/ \rm{s} / \rm{Mpc}$ when applied to the low-$z$ Pantheon+ dataset. Despite the fact that we have only used the diagonal part of the covariance matrix, since this type of correction is systematic and of cosmological origin we expect a similar shift to occur also when the correlations between different uncertainties are accounted for. It is interesting to note that the SH0ES collaboration also find a larger value of $H_0$ of $\approx 0.3 ~\rm{km}/ \rm{s} / \rm{Mpc} $ when only SN~Ia in the range $0.06>z>0.0.15$ are considered in the analysis (see variant 35 in Table 5 of Ref.~\cite{Riess:2021jrx}). This corroborates our results, showing that removing the seemingly most biased sources (essentially those within Laniakea) leads to larger values of the Hubble constant.\footnote{We are grateful to the anonymous referee for pointing us to this interesting result. }

\subsection{Surface Brightness Fluctuations and impact on $H_0$}

\begin{figure}[h]
  \includegraphics[scale=0.7]{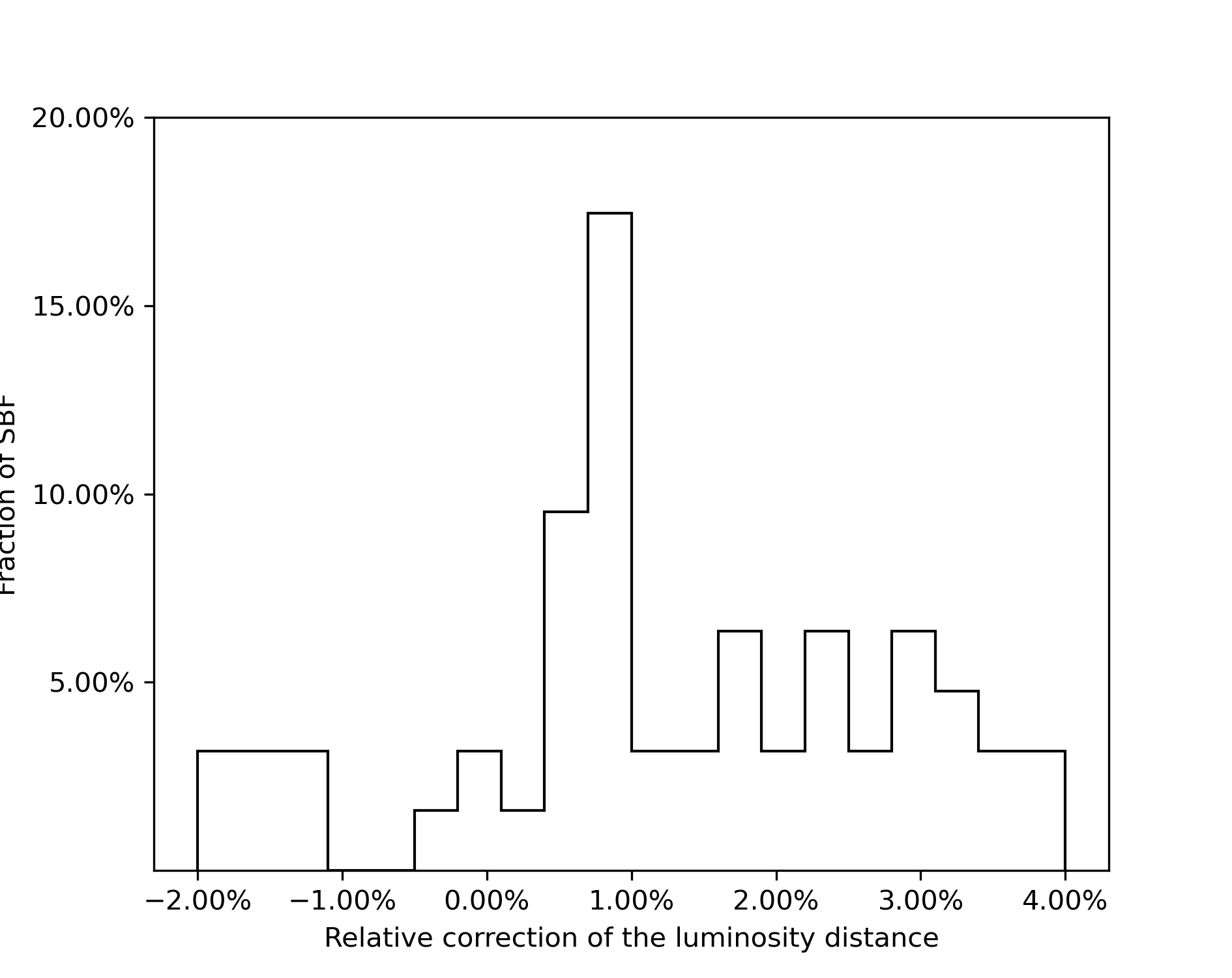}\\
	\caption{ The distribution of the corrections for the SBF sources from Ref.~\cite{2021ApJS..255...21J}}\label{SBFcorr-radec-hist}
\end{figure}

\begin{figure}[h]

    \includegraphics[scale=0.78]{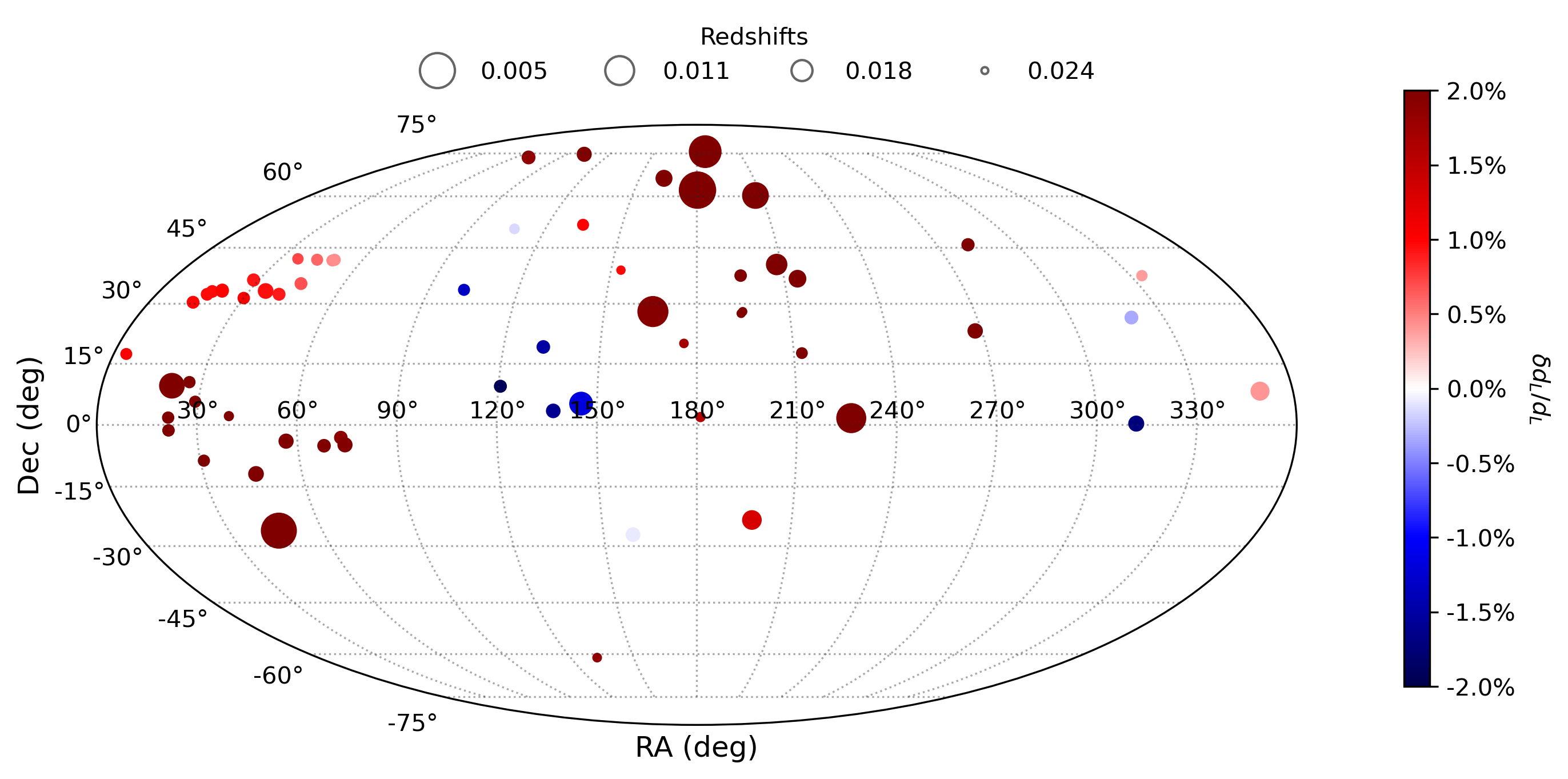}
	\caption{The position on the sky of SBF sources from Ref.~\cite{2021ApJS..255...21J}, color-coded with the amplitude of the relative correction induced on them by Laniakea's backreaction.}\label{SBFcorr-radec}
\end{figure}

Surface Brightness Fluctuations (SBF) are a promising cosmological tool providing an alternative way to calibrate the intrinsic magnitude $M$ of SN~Ia, which can also be used to infer directly the value of $H_0$. It is interesting to compare the impact of our corrections on these sources with the one they had on Pantheon+, to get a qualitative estimate of how Laniakea's backreaction affects a less sparse distribution of data at lower $z$. 
We use SBF measurements from Ref.~\cite{2021ApJS..255...21J}, which consist of a catalog of 63 sources with corresponding observed distance modulus, uncertainty and velocities (from which we infer the redshifts). In Fig.~\ref{SBFcorr-radec-hist} we plot the histogram of these corrections, and in Fig.~\ref{SBFcorr-radec} the positions of the sources in RA and Dec, color-coded with the corresponding corrections. Being on average closer to the observer than the low-$z$ sources in Pantheon+, the average correction is bigger. Using again the low-$z$ cosmographic expansion of $d_L$ and minimizing the sum of the residuals we find a shift in the best fit values $\Delta H_0= H_0^{\rm{Lan}} -\bar{H}_0 \simeq 1.4 \;\rm{km}/\rm{s}/\rm{Mpc}$.

\section{Discussion}\label{conclusion}
The interplay between the constituents of the cosmic web and the coarse-grained description endorsed by the Copernican principle remains an elusive riddle of modern Cosmology. This work tries to shed some light on the subject by deriving an effective description of the weak backreaction of our local Universe on Cosmological observables. We focus on the Laniakea supercluster, the gravitational basin of attraction hosting the Milky Way. The rationale behind the choice of Laniakea's boundary as averaging scale is readily explained: by definition of gravitational basin, geodesics of test masses (i.e. streamlines) starting in its interior are confined within the latter,  thereby remaining ``isolated'' from the rest of the Universe (apart from its the background evolution). 

In order to capture Laniakea's spatial anisotropies we propose an ellipsoidal model exhibiting triaxial expansion. To keep it simple, we neglect the time evolution of the anisotropies and assume a constant expansion rate along each direction. Furthermore, we approximate its interior as homogeneous and spatially flat. Figs.~\ref{ResDistSpace} and \ref{Homtest} show the goodness of the latter assumptions within the precision of current data. The different expansion rates inside and outside the ellipsoid induce a line-of-sight dependent correction to the comoving distances computed in a FLRW background, cf. Eq.~\eqref{deltachi}. We estimate the impact of these corrections on luminosity distances in Eq.\eqref{deltad}, and plot in Fig.~\ref{corrfullsky} their values as a function of RA and Dec for an idealized distribution of sources isotropically distributed in the redshift range $0< z< 0.1$, as seen by an observer in the Milky Way. 

To assess their impact on cosmological inference, we computed these corrections for two low-$z$ catalogs of standardizable sources used to measure the value of $H_0$, SN~Ia from Pantheon+ \cite{Scolnic:2021amr,Riess:2021jrx} and SBF of early-type massive galaxies from Ref.~\cite{2021ApJS..255...21J}. A visual summary of these corrections and their distribution on the sky is given in Figs.~\ref{Pantheon+_corr} and \ref{SBFcorr-radec}. Using the cosmographic low-$z$ expansion of the luminosity distances \cite{Visser:2004bf} corrected with Laniakea's backreaction, and adopting a reference value of $q_0=-0.51$ for the deceleration parameter, we found that the best fit values for $H_0$ are shifted towards higher values of $\Delta H_0^{\rm{SN~Ia}}\approx 0.6 \;\rm{km}/\rm{s}/\rm{Mpc}$ and $\Delta H_0^{\rm{SBF}}\approx 1.4 \;\rm{km}/\rm{s}/\rm{Mpc}$ respectively. The latter is significantly bigger than the former because of the redshift distribution of the sources. SBF, being observed at lower redshifts, will be more affected by Laniakea's backreaction as the latter is inversely proportional to the comoving distance to the source.
From a fundamental point of view these shifts are not surprising. As mentioned in section \ref{Laniakea}, Laniakea is on average overdense and thus the expansion rate inside, on average, slower. For an apparent source outside Laniakea to have the same apparent magnitude $m$ as in a FLRW background, therefore, it is required that the expansion rate outside is on average larger. In other words, the value of the Hubble rate inferred from a source outside the structure will be akin to a weighted average between the two different expansion rates, with the weights depending on the ratio between the lengths of the paths inside and outside Laniakea.  

These results seemingly worsen the recently established tension between the inferred value of $H_0$ from early and late times Universe probes, which has been argued to potentially be the sign of new cosmological physics (see for example Refs.~\cite{Vagnozzi:2019ezj,Camarena:2023rsd,Abdalla:2022yfr,Perivolaropoulos:2021jda,DiValentino:2021izs,Hu:2023jqc}). This might appear to be in contradiction with the possibility, explored for example in Refs.~\cite{Hoscheit:2018nfl,Shanks:2018rka,Ding:2019mmw,Cai:2021wgv,Perivolaropoulos:2023tdt,Alestas:2020zol,Perivolaropoulos:2023iqj,Marra:2021fvf,Alestas:2021nmi,Perivolaropoulos:2021bds,Desmond:2020wep}, that local gravitational physics could alleviate the Hubble tension. Amongst these, a class of models achieve a lowering of $H_0$ under the assumption that we live in an underdense region, whose inner expansion rate is on average larger than the background one. Some results in the literature, see for example  refs. \cite{Bohringer:2019tyj,Whitbourn:2013mwa,Tokutake:2017zqf}, seem to corroborate the latter assumption finding evidence of local voids which averaged on spheres of $ r\gtrsim 100 \rm{Mpc}$ have density contrasts of $\delta\leq-0.1$, unexpected within the $\Lambda$CDM model. Computing the average density contrast of a sphere centered in Laniakea with radius $r\approx 110 \;\rm{Mpc}$ (i.e. the average distance of the boundary of the ellipsoid from the center) using the CF4 reconstruction we found $\delta\sim-0.06$, within the prediction of the concordance model (see for example Fig. 6 of Ref.~\cite{Camarena:2022iae}), and in agreement with the expectations from the Sibelius-DARK constrained realisation simulation (see Fig.4 of Ref.~\cite{McAlpine:2022art}) and sample variance in the Supernovae distribution \cite{Zhai:2022zif}. However, this sphere is not centered in the Milky Way, which might explain why the result differs from the aforementioned ones. Indeed, overdensities such as Laniakea are surrounded by voids (from which they have collected matter), and therefore any sufficiently large spherical average will include these under-dense regions. On the other hand,  Refs.~\cite{Camarena:2022iae,Camarena:2023rsd,Cai:2020tpy} also found no evidence of any large void or overdensity, thus disfavoring a local resolution of the Hubble tension. Our analysis corroborates these results, suggesting instead that the tension is likely to be (slightly) worsened by Laniakea's backreaction.   
An important caveat, however, is that our analysis does not exclude the possibility that large voids in the annular region between $110-400 \;\rm{Mpc}$ outside Laniakea could balance and overcome the backreaction from Laniakea, like a rather picturesque Matryoshka doll. Alternative modelling choices accounting for the impact of these voids are therefore required to fully understand the impact our cosmic environment's gravitational backreaction, which will be the focus of forthcoming studies.\footnote{After the publication of this work, follow up analyses of inhomogeneous modeling of our local environment have been carried in Refs.~\cite{Galoppo:2025hzy,Sorrenti:2026tye}, both in excellent agreement with our results.}   

Whilst our results suggest that Laniakea's gravitational backreaction cannot provide a resolution to the current cosmological tensions, its understanding is crucial to assess whether the interplay between different mechanisms can satisfactorily handle the problem, as suggested recently in Ref.~\cite{Vagnozzi:2023nrq}, and to understand the variations observed in the determination of the Hubble constant from different sampling of sources at low and high redshifts, see for example Refs.\cite{Krishnan:2021jmh, Yu:2022wvg,McConville:2023xav}.  In particular, Laniakea's neighbours and the voids between them might play an important role in the resolution of the Hubble tension, which motivates for better mapping and modelling of our cosmic environment. 
To assess the impact of the local Universe on our cosmological inference one needs two ingredients; an averaging scheme and an averaging volume (scale). With no fundamental reason to determine either of the two, in this work we suggested a data-driven approach based on a sensible definition of large scale structure. Whilst the choice of a simplistic ellipsoidal shape for Laniakea might not be the most rigorous one, it is a first step to capture the impact of its anisotropies. Upcoming surveys like DESI \cite{Saulder:2023oqm} and 4HS \cite{4HS} are expected to measure with unprecedented accuracy peculiar velocities up to redshift $z\approx 0.15$, and in the next future we will have in our hands extremely detailed maps of the Universe around us. In preparation for that, it is important to learn how to navigate.     

\subsection*{Acknowledgments}
We are grateful to Asta Heinesen, Valerio Marra, David Camarena and Brent Tully for their useful comments and suggestions. We are also grateful to Dr Boris Kriger for identifying a few errors in our publicly available code and bringing them to our attention.
\noindent L.G., C.H., K.S., and T.D. acknowledge support from the Australian Government through the Australian Research Council Laureate Fellowship grant FL180100168. S.V. acknowledges support from the Istituto Nazionale di Fisica Nucleare (INFN) through Commissione Scientifica Nazionale 4 (CSN4) Iniziativa Specifica ``FieLds And Gravity'' (FLAG), and from the University of Trento and the Provincia Autonoma di Trento (PAT, Autonomous Province of Trento) through the UniTrento Internal Call for Research 2023 grant ``Searching for Dark Energy off the beaten track'' (DARKTRACK, grant agreement no.\ E63C22000500003). This publication is based upon work from the COST Action CA21136 ``Addressing observational tensions in cosmology with systematics and fundamental physics'' (CosmoVerse), supported by COST (European Cooperation in Science and Technology).”

\appendix
\section{Comparing luminosity distances in FLRW and our Toy Model}\label{appA}
In a pure FLRW Universe the redshift of a photon emitted at the emission time $t_e$ observed at $t_0$ is given by:
\begin{equation}
1+\bar{z} = \frac{a(t_0)}{a(t_e)}\;.    
\end{equation}
In a Bianchi I Universe, a photon emitted at the same time $t_e$ will have instead a different redshift given by Eq.~\eqref{redBI}.
To simplify the notation, let us refer to those quantities computed using FRLW redshifts with a barred subscript, e.g. $\bar{\chi}(\bar{z}_A)\equiv\bar{\chi}_{\bar{A}}$. Notice that this is in general different from $\chi_A(\bar{z}_A)\equiv \chi_{\bar{A}}$. Using this notation, remembering that the expansion rate outside Laniakea is the same in the two models, we can write the difference between Eqs. \eqref{dflrw} and \eqref{dlan} for sources at $z_B$ and $\bar{z}_B$ respectively as
\begin{equation}\label{d-dcomp}
    d_B - \bar{d}_{\bar{B}} = \left(1+\bar{z}_B \right)\left[\left(\chi_A-\bar{\chi}_{\bar{A}}\right) + \left(\bar{\chi}_{AB} -\bar{\chi}_{\bar{A}\bar{B}}\right)\right] + \epsilon\left(\chi_A + \bar{\chi}_{AB} \right)\;,
\end{equation}
where we have introduced $\epsilon\equiv z_B -\bar{z}_B$. Let us compute explicitly the various terms entering the above expression for fixed $\theta,\phi$. It is helpful to also define the quantity $\eta\equiv z_A - \Bar{z}_A$, which is the difference in the redshift which an observer would measure for photons emitted from a source at the boundary edge of Laniakea at a time $t_A$ in models with or without inhomogeneity. Assuming that $\epsilon$, $\eta$ and $\Delta H/\bar{H}$ are small quantities, and that $\Delta H$ along each line-of-sight is constant we can write: 
\begin{equation}\label{firstterm}
\begin{split}
\left(\chi_A -\bar{\chi}_{\bar{A}}\right)=& c\left(\int_0^{\bar{z}_A + \eta}d\tilde{z}\; \frac{1}{H(\tilde{z})} -\int_0^{\bar{z}_A}d\tilde{z}\; \frac{1}{\bar{H}(\tilde{z})} \right)\\
 =& c\left(\int_0^{\bar{z}_A}d\tilde{z}\; \left(\frac{1}{H(\tilde{z})}-\frac{1}{\bar{H}(\tilde{z})}\right) + \int_{\bar{z}_A}^{\bar{z}_A + \eta}d\tilde{z}\frac{1}{H(\tilde{z})}\right)\\
 \approx& \; \Delta \chi_{A} + c\frac{\eta}{H(\bar{z}_A)}\;, 
\end{split}
\end{equation}
where $\Delta \chi_A$ has been defined in Eq.~\eqref{deltachi}, and
\begin{equation}\label{secondterm}
\begin{split}
    \left(\bar{\chi}_{AB}-\bar{\chi}_{\bar{A}\bar{B}}\right)&= c\int_{\bar{z}_A+\eta}^{\bar{z}_B+\epsilon} d\tilde{z} \frac{1}{\bar{H}(\tilde{z})} - c\int_{\bar{z}_A}^{\bar{z}_B} d\tilde{z} \frac{1}{\bar{H}(\tilde{z})}\\
    &=c\int_{\bar{z}_B}^{\bar{z}_B+\epsilon} d\tilde{z} \frac{1}{\bar{H}(\tilde{z})} - c\int_{\bar{z}_A}^{\bar{z}_A +\eta} d\tilde{z} \frac{1}{\bar{H}(\tilde{z})}\\
    &\approx\; c\left(\frac{\epsilon}{\bar{H}(\bar{z}_B)} -\frac{\eta}{\bar{H}(\bar{z}_A)} \right)\;,
\end{split}    
\end{equation}

\begin{equation}\label{thirdterm}
\begin{split}
 \left(\chi_A + \bar{\chi}_{AB}\right) &= c\left(\int_0^{z_A}d\tilde{z} \frac{1}{H(\tilde{z})} + \int_{z_A}^{z_B}d\tilde{z} \frac{1}{\bar{H}(\tilde{z})} \right)\\
 &= c\left(\int_0^{z_A}d\tilde{z} \frac{1}{\bar{H}(\tilde{z}) + \Delta H} + \int_{z_A}^{z_B}d\tilde{z} \frac{1}{\bar{H}(\tilde{z})} \right)\\
 &\approx \; c\left(\int_0^{z_B}d\tilde{z} \frac{1}{\bar{H}(\tilde{z})} - \int_0^{z_A}\frac{\Delta H}{\bar{H}^2(\tilde{z})}\right)\\
 &\approx \bar{\chi}_{\bar{B}} + c\left(\frac{\epsilon}{\bar{H}(\bar{z}_B)} - \Delta H\int_0^{z_A}d\tilde{z} \frac{1}{\bar{H}^2(\tilde{z})}\right)\;.
\end{split}
\end{equation}
Combining Eqs.~\eqref{firstterm},\eqref{secondterm} and \eqref{thirdterm} we can rewrite Eq.~\eqref{d-dcomp} as
\begin{equation}
    d_B -\bar{d}_{\bar{B}} = \left(1+\bar{z}_B\right)\left[\Delta \chi_{A} +c\left(\frac{\epsilon}{\bar{H}(\bar{z}_B)} + \frac{\eta}{H(\bar{z}_A)} - \frac{\eta}{\bar{H}(\bar{z}_A)}\right) \right] + \epsilon\left[\bar{\chi}_{\bar{B}} + c\left(\frac{\epsilon}{\bar{H}(\bar{z}_B)} - \Delta H\int_0^{z_A}d\tilde{z} \frac{1}{\bar{H}^2(\tilde{z})} \right)\right]\;.
\end{equation}
Keeping only linear order terms in $\eta$,$\epsilon$ and $\Delta H/H$ and dividing both sides of this equation for $\bar{d}_{\bar{B}}$ we can finally write the relative difference of the luminosity distances as:
\begin{equation}\label{dcomparison}
   \frac{d_B -\bar{d}_{\bar{B}}}{\bar{d}_{\bar{B}}} \approx \frac{\Delta \chi_A}{\bar{\chi}_{\bar{B}} } + \frac{\epsilon}{1+\bar{z}_B}\;. 
\end{equation}
Interestingly, the above expression shows that differences due to $\eta\neq 0$ only appear at second order. Eq.~\eqref{dcomparison} has been obtained comparing the luminosity distance to the ``same source'' $B$ in the two models, where with ``same source'' we refer to a photon emitted at the same cosmological time $t_e$ along the same line-of-sight. In practical applications, however, one can only observe redshifts, which is the reason why in the main text, Sec.~\ref{distances_comparison}, we identify the ``same source'' with one having the same observed redshift $z_B$. It is straightforward to realize that Eq.~\eqref{dcomparison} reproduces Eq.\eqref{deltad} once we set $\bar{z}_B = z_B$, or equivalently $\epsilon=0$.
\begin{table}[h]
\centering
\begin{tabular}{c|c|c|c}
\hline
Variable & meaning & Variants & meaning \\
\hline
$v_{\rm BF}$ & bulk flow velocity inside Laniakea & $v_{\rm BF}^x, v_{\rm BF}^y, v_{\rm BF}^z$ & Components of BF in supergalactic frame \\
$v_r^{\rm rcst}$  & Measured (reconstructed) {\rm radial} pec vel &  &  \\
$v_{\rm pec}$ & l.o.s.\ peculiar velocity of source &  &  \\
$v_{o}$ & l.o.s.\ peculiar velocity of observer &  &  \\
$v_{\rm S}$ & vector peculiar velocity of source & $v_x$, $v_y$, $v_z$ & components of vector PV \\
$v_{\tilde{r}}^{\rm th}$ & l.o.s.\ peculiar velocity predicted by the toy model & $v_{\tilde{r}}^{n, {\rm th}}$ & indexed version\\
$\bar{\chi}$ & Comoving distance in FLRW & $\chi$ & Comoving distance in Laniakea model \\
$\bar{H}$ & Hubble parameter in FLRW & $H, H_i $ & Hubble paramrameter inside Laniakea\\
$H_{\rm{los}}$ & Directional Hubble parameter  &  & \\
$\Delta H_i$ & $H_i-\bar{H}$ & $\Delta H_a$, $\Delta H_b$, $\Delta H_c$ & Components of $\Delta H_i$\\
$\Delta H_{\rm sph}$ & Hubble factor for a spherical model & $\chi^2_{\nu\; \rm{sph} } $ & Reduced $\chi^2$ for spherical model \\
$\bar{d}_L$ & Luminosity distance in FLRW & $d_L$ & Luminosity distance in Laniakea model \\
$\bar{d}_A$ & Ang.\ diam distance in FLRW & $d_A$ & Ang.\ diam  distance in Laniakea model \\
$z_{\rm obs},\;z_B$ & Observed redshift and observed redshift for a source $B$ & $\bar{z}$& redshift in a pure FLRW Universe\\

\hline
\end{tabular}
\caption{Summary of symbols used in the paper and their meaning.}
\label{listofvariable}
\end{table}

\bibliography{Laniakegg.bib}

\end{document}